\input amstex
\documentstyle{amsppt}
\input epsf.tex


\catcode`\@=11
\def\logo@{}
\catcode`\@=\active

\def\isom{\cong}

\pageheight{46pc}
\def\epi{\twoheadrightarrow}

\mag\magstep 1
\NoBlackBoxes\tolerance10000

\def\ss{\smallskip}

\def\<{\langle}
\def\>{\rangle}

\def\sk#1{^{(#1)}}
\def\inv{^{-1}}

\TagsAsMath
\def\ni{\noindent}
\def\Z{\Bbb Z}
\def\R{\Bbb R}

\def\Q{\Bbb Q}
\def\a{\alpha}

\def\bb{\beta}
\def\g{\gamma}
\def\d{\delta}  

\def\z{\zeta}

\def\l{\lambda}

\def\k{\kappa}

\def\C{\Bbb C}

\def\Aut{\hbox{\rm Aut}}

\def\D{\Delta}



\def\({\hbox{\rm(}}
\def\){\hbox{\rm)}}

\input epsf.tex
\input amstex
\documentstyle{amsppt}
\input epsf.tex


\catcode`\@=11
\def\logo@{}
\catcode`\@=\active

\pageheight{46pc}
\def\isom{\cong}
\def\epi{\twoheadrightarrow}

\mag\magstep 1
\NoBlackBoxes\tolerance10000

\def\ss{\smallskip}

\def\<{\langle}
\def\>{\rangle}

\def\sk#1{^{(#1)}}
\def\inv{^{-1}}

\TagsAsMath
\def\ni{\noindent}
\def\Z{\Bbb Z}
\def\R{\Bbb R}

\def\Q{\Bbb Q}
\def\a{\alpha}

\def\bb{\beta}
\def\g{\gamma}
\def\d{\delta}  

\def\z{\zeta}

\def\l{\lambda}

\def\k{\kappa}

\def\C{\Bbb C}

\def\Aut{\hbox{\rm Aut}}

\def\D{\Delta}



\def\({\hbox{\rm(}}
\def\){\hbox{\rm)}}

\topmatter
\title{Topology of the Standard Model I: Fermions}
\endtitle
\author{Steve Gersten}
\endauthor

\abstract{
The Harari-Shupe model for fermions is extended to a
 topological model which contains an explanation for
  the observed fact that there are only three generations 
  of fermions. Topological explanations are given for
   $\beta$-decay and for proton decay predicted
    in supersymmetry and string theories. 
    An explanation is given for the observed
     fact that the three generations of fermions have 
     such similar properties. The concept of ``color" is
      incorporated into the model in a topologically meaningful way.
       Conservation laws are defined and discussed 
       in the context of the algebraic topology of 
       the model, and preon number is proved to be
        linearly determined by charge, weak isospin, and color.}

\endabstract

\TagsAsMath
\address{Mathematics Department,
University of Utah, 
155 S. 1400 E., Salt Lake City, UT 84112-0090,
{\rm http://www.math.utah.edu/$\sim\null $sg}
}
\endaddress
\email{sg at math dot utah dot edu}
\endemail
\subjclass{81P05, 20F06}\endsubjclass
\keywords{fermion, standard model, Harari-Shupe model, 
generation problem, van Kampen diagram, universal covering}
\endkeywords
\thanks{Supported by Social Security and Medicare}
\endthanks
\thanks{\copyright S. Gersten 2012}
\endthanks
\thanks{I wish to thank Domingo Toledo
for discussions.
I am very grateful to Yong-Shi Wu for explaining
the physics underlying ``color".
Wikipedia articles on the standard model and
related topics were useful and my thanks go
to their anonymous authors .}
\endthanks

\endtopmatter

\document

\def\00{[0]}
\def\11{[1]}
\def\22{[2]}

\heading{\bf Preface to help non-grouptheorists get a start
in reading this paper}
\footnote{There is something new for grouptheorists here as well.  In the 
word problem all that matters is the minimum number of 2-cells in a van Kampen
diagram filling a relation~$R$ of a presentation, called the area of $R$.  The
diagram itself is irrelevant for the word problem.  What is new in these
articles is that the van Kampen diagrams are given an independent
physical meaning.  Thus both the proton and the positron are fillings of the
word $aaa$ in the presentation $P$ to be defined shortly.  But they have 
very different topological and physical properties.
}
\endheading

Since some readers may not be familiar with the term van Kampen diagrams,
I can refer them to an excellent article
in Wikipedia
\footnote{
  http://en.wikipedia.org/wiki/Van$\underscore{\null}$Kampen$\underscore{\null}$diagram}; I'll give a brief explanation shortly.  The model
I propose for particles of SM (standard model of particle physics)
makes essential use of them, but the idea can be understood
at a non-technical level as follows.

The model for particles 
is like Buckminster Fuller's geodesic dome.  The particles are
the skin spanning the skeletal frame and the preons [Li] are the
segments of that frame. So {\bf particles are 2-dimensional and 
preons are 1-dimensional.\/}  It makes no sense to say a particle
(like a quark or electron) is composed of preons, in the way a sack of
gum drops is composed of pieces of candy.  A particle is the
surface spanned by the 3 segments of Fuller's frame that bound it.
These segments have names, $a$ or $b$, and orientations,
that is, arrows.
\footnote{Figure~2 of this article is an example of
a van Kampen diagram that represents a particle, the proton in
this case, with the labels $a,b$ on the edges.}
 It will turn out that, for all particles in our theory,
the arrows on one spanning triangle all point in the same direction;
this is a deep fact and the explanation will be delayed to the
sequel paper on bosons, but it is connected with the intrinsic
spin of a particle.
Particles may be detected, albeit indirectly, as in the case of
quarks, which are not directly observable (each quark has a color
and only ``colorless" combinations of them (\S 4 below) are observable),
but the segments of the skeleton 
may never
be seen, but only inferred by the characteristics of surfaces spanning them.
\footnote{The usual interpretation of
the Harari-Shupe model is like the skeleton of a {\it fish\/}, whereas mine
is like that of a {\it lobster\/}.  Theirs has the skeleton on the inside,
whereas I've moved it to the outside.
}

I want to state from the beginning that I have only changed slightly the
interpretation of the basic idea of [Ha] and [Sh], but not the
idea itself.  By interpreting their
preons 
\footnote{The terminology ``preon" is used in [Li] but does not
occur in [Ha] nor in
[Sh], which use the terms ``rishon" and ``quip" respectively.}
as the framework (or skeleton) rather than the content of
particles, I can solve the problems that these papers pose
concerning the standard model.  It's the paper I would have expected
either of these authors to write had they been aware of 
the (admittedly arcane) field of combinatorial group theory.
Thus my goals in this article are only to go as far as these authors could have
gone in 1979 with the reinterpretation of ``preons".  In the sequel
article, I shall take the idea a step further by incorporating spin.

Now for a quick definition of van Kampen diagrams.  It is
the primary tool of combinatorial group theory (also known as
geometric group theory).  This branch of group theory is the
study by geometry and topology of group presentations.
A group presentation consists of generators and defining relations.
A van Kampen diagram is a geometric way of representing
relations.  For more detail and for some superb drawings
of van Kampen diagrams, consult the Wikipedia article.

An example of a presentation is that whose generators are
all the edges of Fuller's geodesic dome and whose defining relations
are those determined by the set of triangles of the dome.  We could call the group presented
Buckminster Fuller's group; it has not been
studied to my knowledge. One van Kampen diagram is the dome itself.
\footnote{http://carlacapeto.files.wordpress.com/2010/11/buckminster-fuller-dome.jpg}

The only presentation we are concerned with in these articles is
$P=<a,b; \ aaa, bbb, aab, abb>$.  The group presented by~$P$ is
the cyclic group of order~3.  
Two van Kampen diagrams in $P$ may be seen in
Figures~1 and~2 of this article.  The second represents a proton and the
three curvilinear triangles are, reading from left to right, the up,
down, and up quarks.
These two diagrams are of particular interest because they
may be considered ``spherical diagrams" by appending a face
at infinity.  So, with the exterior of the diagram compactified by adding
a point at infinity, it becomes a sphere divided into four curvilinear
triangles.  These two spherical diagrams play a central role in
characterizing SM in \S 5 below.
\footnote{I have adopted Domingo Toledo's suggestions
to revise
an earlier attempt at this preface in order to make it
less technical.}

I cannot emphasize strongly enough that what is {\bf{not}\/} being attempted
is to calculate masses of particles nor to compute the fundamental constants
of nature.  The topological approach cannot and does not attempt to do this.

All that
topology can do is to establish limits of what is possible, but it cannot
fill in the details.  For example my theory tells us that there
is no fourth generation of fermions, but it cannot fill in the details
about the masses of the three generations of fermions that exist.  Only
a metric theory can attempt to do this, and the mere beginnings of a metric
theory are sketched at the end of the second paper on bosons.
Do not expect me to calculate the value of ``$g$" anytime soon.

Professor Wu suggested that I should emphasize also that the
complex~$K$ and its universal cover~$\tilde K$,
in which all my constructions take place, do not lie in
3-space (or even 4-space), but should be considered in a larger
dimensional space, possibly in the fibre of a fibre bundle over Minkowski space.  He also pointed out that the discreteness
of the complex~$K$ (as opposed to a continuous geometry)
was a good feature that might sweeten the pill of
higher dimensions.  Those physicists who accept higher dimensional
spaces as the venue of physics will appreciate this while those
who reject them will reject this article whatever I do to try to 
reach them.

\ss
\ss

\centerline{\bf Contents}
\ss
\ss

Preface to help non-grouptheorists get a start in what is going on

\S 0. Introduction

\S 1. Review of Harari-Shupe model

\S 2. Conjugacy classes in a free group

\S 3. 2-cycles in the 2-complex $K$

\S 4. Color

\S 5. Standard Model

\S 6. Neutrino Oscillations

\S 7. Questions and comments

Appendix 1. Historical remark

Appendix 2. Color group $M$ and $A_2$

Appendix 3. Conservation laws in SM and quantum numbers

Appendix 4. $\beta$-decay revisited

Appendix 5. $\Delta^{++}$ baryon

Appendix 6. Quantum numbers and the first Chern class

Appendix 7. Colorings as conservation laws

Appendix 8. Preons are determined by charge, color, and weak
isospin

Appendix 9. Dark Matter

Appendix 10. Nucleogenesis

Appendix 11.  The dual picture
and $\Delta^{++}$ test 

Appendix 12.  Entanglement

Disclaimer.

\heading{\bf \S 0. Introduction}
\endheading
In this article we propose a topological model for SM, the standard model
of particle physics.  
 A particle will turn out to be represented by a class
of colored van Kampen diagrams in the universal cover $\tilde K$ of a 
complex $K$ we construct in \S 3
 (diagrams related by diamond moves represent the
same particle; the concept of color is introduced in \S 4 below).  The boundary
label of a diagram is a conjugacy class $\frak c$ of a single element 
in $F=F(a,b)$, the free group with free basis $\{a,b\}$, and is thus determined
by the particle.  Its antiparticle is represented by that diagram with the
opposite orientation, and hence its boundary label is $\bar \frak c$,
the conjugacy class of inverses of elements of $\frak c$.  If
a particle is its own antiparticle (as is the case for the photon, gluons, and the
$Z$-boson), then $\frak c=\bar\frak c$, which implies that $\frak c$
consists only of the neutral element~$1$ of $F$.  Not all diagrams represent
observable particles; the observability criterion is given in \S 4 below.
At the end of \S 3 we illustrate the notions with
 van Kampen diagrams representing the
proton, neutron, and their antiparticles without indicating the
colors that must
be attached to the faces.

The notions are unified in \S 5 to give a topological formulation for
SM, including predictions for all particles, observable or not, which
can exist.  The emphasis in this paper is on fermions, which are the
easiest to characterize, but the model applies to bosons as well.  The
deduction of the various species of bosons will be left to a future
paper, but we have included a diagram for a Higgs boson, since 
people I have spoken to about this article have requested to see it
and it is in the news because of its discovery by the LHC 
at CERN last year.

Finally, in \S 6, a purely topological explanation
 is given based on our model for
the neutrino deficit, the fact that the neutrino
flux from the sun observed in the laboratory is one third
of the rate predicted by theory.

The correspondence between particles and van Kampen diagrams is the basic
idea of this article, which is an expansion of ideas in papers of 
Harari [Ha] and Shupe [Sh] and which have been recently revived in an
expository artice [Li].  Viewing particles in this way leads to the
complete model in \S 5, including answers to several of the questions 
raised in these papers.  Among these are why there are only three
generations of fermion, why the generations have such similar properties, and
where the prediction of new fundamental particles stops.

A general reference for the algebraic topology used is
[AH] and a reference for notions of combinatorial group
theory is [BRS].  Other references to web articles will
be provided as needed in footnotes.

There are a number of appendices which examine finer
point of the theory.  Appendix 12
offers a geometrical explanation for the paradoxes of quantum
entanglement.  In fact geometry is the overall theme of the
paper, that fundamental physical facts can be explained by
geometry even when the numbers themselves have to be
determined by experiment.  Appendix 11 shows how to replace
the notion of van Kampen diagram, which offended at least one
reader (see the letter quoted in the Disclaimer), by a planar graph with additional labels.  This is
an equivalent formulation, but the van Kampen formulation is
still needed because covering space theory applies directly to it,
whereas the graph formulation is more intuitive for physicists.

\heading{\bf \S 1.  Review of Harari-Shupe model}
\endheading

Harari [Ha] and Shupe [Sh]  have proposed a model wherein quarks 
and electrons are
composite particles, composed each of three fundamental particles
called ``preons" by Lincoln in a recent article [Li].  In my
notation the fundamental particles are $a$ and $b$ and their
antiparticles are $\bar a$ and $\bar b$, respectively.
\footnote{Both Shupe and
Lincoln call $a,\bar a, b, \bar b$ by $+, -, 0, \bar 0$ respectively,
whereas Harari calls them $T,\bar T, V, \bar V$.
}
The first generation of fermions
consists of the
 electron $e$, up quark $u$, down quark $d$, and electron neutrino $\nu_e$;
 these
are  $\bar a\bar a\bar a, aab, \bar a \bar b\bar b$ and
$bbb$, and their antiparticles are $aaa, \bar a\bar a\bar b,
abb$, and $\bar b\bar b\bar b$, respectively.
Then the proton~$p$ is consists of two up and one down quark,
or $p=aab +aab+\bar a\bar b\bar b$, whereas the neutron~$n$
is two down and one up quark, or $n=\bar a \bar b\bar b+
\bar a\bar b\bar b+aab$.  When one assigns an electric charge
of $+\frac 1 3$ to $a$ and $0$ to $b$ and negatives of these
numbers to the antiparticles, then the charges add up to
the correct numbers~$1$ for the proton and $0$ for the neutron.

There are two distinguished decay processes identified by Harari,
$d\to u+e+\bar\nu_e$ and $u+u\to \bar d+e^+$, where $e^+$ is the
positron, antiparticle to the electron, 
and $\bar\nu_e$ is the electron antineutrino.
In my notation these are 
$\bar a\bar b\bar b\to aab+\bar a\bar a\bar a+\bar b\bar b\bar b$
and
$aab +aab\to abb+aaa$.  
The first of these processes underlies $\beta$-decay or
$n\to p +e+\bar \nu_e$ and the second underlies the decay of a 
proton which is predicted by supersymmetry and string theories.
A hypothesis made is that the net number of $a$'s and of $b$'s
(that is, the number of $a$'s minus the number of $\bar a$'s
and similarly for the $b$'s) on one side of the reaction must
equal the net number individually on the other side.  So there 
is conservation of the net preon number of $a$'s and of $b$'s.

It should be emphasized that [Ha] and [Sh] were published in 1979
and there has been no confirmation of the prediction that these
fermions are composite particles to date.  In addition, there
are two further generations of fermions identified in the
laboratory, where the second generation is
the muon, charm and strange quarks, and muon neutrino,
and the third generation is the tau particle, top and bottom
quarks, and tau neutrino.  These appear to have similar properties
to the first generation except for significantly higher masses.
One of the most important open problems in particle physics to
date has been to
explain why there appear to be only three generations of fermions
and why the three generations so closely resemble each other.

The reason this thirty year old puzzle is of interest today, and
the reason [Li] was written, is that the large hadron collider
LHC has come into operation and has discovered the existence of 
what appears to be a Higgs boson, the particle predicted over 40
years ago which is necessary to complete the ``standard model" SM
of particle physics.  When the LHC is upgraded next year, it is
hoped that other puzzles will be similarly unraveled, among which 
is the composite nature of fermions.


\heading{\bf \S 2. Conjugacy classes in a free group
}
\endheading

There is a product among conjugacy classes $\frak c, \frak d$ 
in a group whose result
$\frak c\frak d$ is the set of products of elements of the two sets.
The product is associative and commutative.  For each conjugacy class
$\frak c$ there is the conjugacy class $\bar\frak c$ which consists
of inverses of elements of $\frak c$.

We shall consider conjugacy classes of single elements in a free group
$F$ with free basis $\{x_1, x_2,\dots, x_n\}$.  Each such conjugacy class
has a representative which is a cyclically reduced word in the free
basis.  Denote the conjugacy class of $w\in F$ by $[w]$.

Given a set of conjugacy classes of words $[w_1], [w_2],\dots,[w_m]$, we
can form the CW complex $K$ whose 1-skeleton is a bouquet of circles
corresponding to the generators~$x_i$ of $F$ and whose 2-cells~$e_{[w_j]}$ 
have attaching maps the words $w_j$ in the 1-skeleton.  We make
the convention that we attach only one 2-cell for the pair of conjugacy
classes~$[w], [\bar w]$ and that the 2-chain they determine satisfies
thus $e_{[w]}=-e_{[\bar w]}$.
\footnote{This convention will play an important role in the sequel
paper on bosons in the algorithm for determining spin of composite
particles.}
Note that 2-cycles $Z_2(K)$ (with integer coefficients) are integer
linear combinations of the $e_{[w_j]}$ such that the net algebraic sum
of each of the $x_i$ in the attaching maps is zero (occurrences of $x_i$
count as +1 and occurrences of ${\bar x_i=x_i}\inv$ count as -1).
  Since each conjugacy class
$[\frak c]$ determines a homology class in $H_1(K)$, this condition is
equivalent to the assertion that the sum of the homology classes determined
by the $[w_i]$ is zero.

\heading{\bf \S 3.  2-cycles in the 2-complex $K$}
\endheading

A {\it klepton\/} is defined as a conjugacy class of one of the
four words $aaa, aab, abb, bbb$
and their inverses
in the free group $F=F(a,b)$ with free basis $\{a,b\}$.
{\it From now on let $K$ be the 2-complex constructed in the preceding paragraph
with 2-cells $e_\k$ corresponding to the four conjugacy classes
$\k=[aaa], [aab], [abb], [bbb]$ in $F$.\/}
 Note
that $e_{\bar \k}=-e_\k$ in the chain group $C_2(K)$ for each
klepton $\k$.

We require a reaction among 
kleptons $\k_1,\k_2, \dots, \k_m$ and $\l_1,\dots, \l_m$,
denoted by $\sum \k_i\to \sum \l_j$,
to conserve algebraic sum of the
number of $a$'s and of $b$'s.
It follows that $\sum_i e_{\k_i}+\sum_j e_{\bar \l_j}$ is a 2-cycle in $K$.
{\it From now on we identify 
the 2-cell $e_\k$ with the fermion corresponding to $\k$.\/}

A key observation is the fact, proved by an easy computation,
that $K$ has fundamental group $\Z/3\Z$, the cyclic group
of order 3.  It follows that each of the 2-cells $e_\k$ of $K$
has 3 lifts to the universal cover $\tilde K$ of $K$.
\footnote{Note that $\tilde K\sk 1$ is the regular covering
space of $K\sk 1$ associated to the kernel of the homomorphism 
$F\to \Z/3Z$ given by $a\to 1, b\to 1 \pmod 3$.}
This gives the 12 fermions of the standard model.  The antiparticles
arise from the observation that each 2-cell $e_\k$ is a map of the
disc into $K$.  So the mapping with the opposite orientation,
corresponding to the chain $e_{\bar \k}=-e_\k$ corresponds to the
antiparticle $e_{\bar \k}$ of the fermion $e_\k$.

\proclaim{Scholium}  There are exactly 3 generations of
fermions in the model $\tilde K$ for the standard model, corresponding
to the lifts of the 2-cells of $K$.  These are the 24 basic
fermions of the standard model. 
\qed
\endproclaim

\proclaim{Proposition}  Every 2-cycle of $K$ is spherical.
\endproclaim

This means that the Hurewicz map $\pi_2(K)\to H_2(K)=Z_2(K)$
is surjective.

\demo{Proof}

Let us assume that
$\a e_{[abb]}+\bb e_{[aab]}+ \g  e_{[aaa]}+\d e_{[bbb]}$ is a 2-cycle,
where $\a,\bb,\g,\d$ are integers.
It follows that we have linear equations
$\a+2\bb+3\g=0, 2\a+\bb+3\d=0$.
From Gaussian elimination it follows that 
$\a=\g-2\d, \bb=-2\g+\d$,
so the general solution is a linear combination
$\g v_1+\d v_2$ with integer coefficients $\g, \d$
of the integral vectors
$v_1=[1,-2,1,0]$ and $v_2=[-2,1,0,1]$.
A second integral basis for the solutions
is 
$v_1+v_2=[-1,-1,1,1]$ and $-v_1=[-1,2,-1,0]$.
It follows that to prove the theorem it suffices
to show that the 2-cycles
$-e_{[abb]}-e_{[aab]}+e_{[aaa]}+e_{[bbb]}$
and $-e_{[abb]}+2e_{[aab]}-e_{[aaa]}$
are spherical 
(observe that these 2-cycles correspond to the
reactions of $\beta$-decay
$\bar a\bar b\bar b\to aab +\bar a\bar a\bar a+\bar b\bar b\bar b$
and 
$aab + aab\to abb + aaa$ identified in [Ha]).
This is accomplished by explicit construction with the
following spherical diagrams.

\vskip .1in
\epsfxsize=3truein
\centerline{\epsfbox{fig1.epsi}}
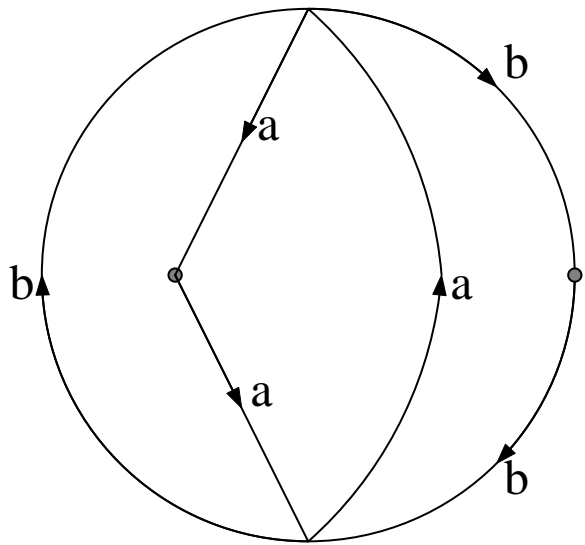
\centerline{Figure 1}

\vskip .1in
\epsfxsize=3truein
\centerline{\epsfbox{fig2.epsi}}
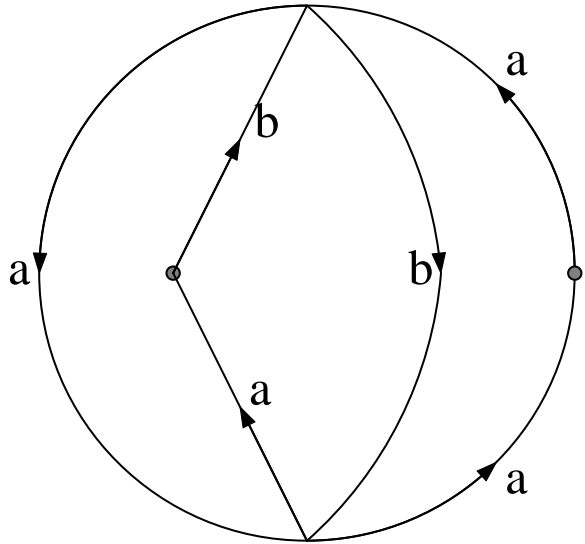
\centerline{Figure 2}

In each of these figures there are four 2-cells, where the fourth
2-cell is at infinity.  One reads the boundary labels for the finite
cells counterclockwise and the boundary label for the cells at infinity
in the clockwise manner.  
Since each 2-cycle is represented by a map of 2-sphere into $K$,
it follows that the generators of $Z_2(K)$ are spherical and hence
all 2-cycles of $K$ are spherical.
This completes the proof.
\footnote
{This result is a special case of the the result that {\it every 
integral 2-cycle on a 2-complex with finite cyclic fundamental
group is spherical\/}. The proof makes use of the edge-term
exact sequence of the Serre spectral sequence along with
the fact that $H_2(G,\Z)=0$ for a finite cyclic group~$G$.
Since we need the explicit form of the generators shown
in Figures~1 and ~2, we have given our direct geometric
argument.
}
\enddemo
\demo{Remark 1}  We shall calculate the kernel of the 
homomorphism $\pi_2(\tilde K)\to H_2(K)$ in \S 6.  It turns 
out to have important physical significance.
\enddemo

The significance of the proposition is the following.  A particle, like
a proton or neutron, is represented by a van
 Kampen diagram $D$ (that is, a combinatorial
map of an oriented singular disc $D$ into $K$) [BRS].  Given a spherical
diagram, one can cut it along one edge to obtain a van Kampen diagram
with freely trivial boundary label.  This can then be combined
with $D$ along an edge with the same boundary label in the same
orientation to obtain a new disc diagram $D'$.  
Then one can do ``diamond moves" ([BRS] page 115) 
on $D'$ and repeat the process
any number of times.  The result of such a sequence of moves
is a reaction among fermions, and we can read off the result 
at the chain level $C_2(K)$ to get the reaction in terms of the
Harari-Shupe model.  The proposition says that all reactions among
fermions, that is, beginning with one collection of fermions and
ending with another, after projecting from $\tilde K$ to
$K$, are obtained in this way.

\demo{Remark 2}  The fact that processes affecting fermions
are represented by spherical diagrams means they lift to the
universal cover $\tilde K$ of $K$.  Thus the same processes
that affect the fermions of the first generation affect those
of the second and third generations.  This explains
why these particles resemble so closely their first generation
counterparts.
\enddemo

To finish this section, we observe that Fig.~2 can be interpreted
as a disc diagram by ignoring the face at infinity.  As such it is
a van Kampen diagram for the proton.
A van Kampen diagram for the antineutron is obtained from it
 by interchanging $a$ and $b$.  Those for the
antiparticles are obtained by reversing all arrows.
\footnote{The other van Kampen diagram for the proton is
obtained by doing a diamond move on the two edges
labelled ``$a$" with origin the bottom vertex. 
}

\heading{\bf \S 4. Color}
\endheading

In this section we take the first steps toward incorporating color,
the source of the strong nuclear force, into our model.  To begin,
let $\00,\11,\22$ denote the residue classes of $0,1,2 \pmod 3$.
Let $N=\00+\11+\22$ in $\Z G$, where $G=\Z/3\Z$
and let $M=\Z G/N \Z G$, where $M$ is defined to be the color group.
In the literature, $\00,\11,\22$ correspond to the colors
red ($r$), green ($g$), and blue ($b$).
Their negatives $-r, -g, -b$ in $M$ are called 
cyan, yellow, and magenta, respectively; they play no
role in the discussion below, but enter in the discussion
of antiparticles and gluons, which we shall not undertake here.
\footnote{In Appendix~2 below I relate my color group
$M$ to the root system~$A_2$.  When I wrote this section
I was ignorant of the existing $SU(3)$ theory and
formulated the notions out of thin air.  After Professor Wu
kindly explained the physicists' theory, I saw
that we were saying the same thing, and the appendix gives
the relation.}
  The designation of colors is
arbitrary and has nothing to do with perceived vision.  It is
important to make the distiction between the color~red $\00$ and
the neutral element $0$ in $M$.  In the literature, $0$ is called
both ``white" and ``colorless".
The discussion below is given in detail for the proton~Fig.~2, but
it can be extended to all the fermions (see the last
paragraph of \S 3).  The gluons will be discussed
in a future paper along with the other bosons of the theory.

We lift the diagram~Fig. 2 for the proton to $\tilde K$.
\footnote{The lift here is a technical
device to facilitate the definition of color.  Technically this means
that color, for example on the quarks constituting a proton, is
defined by the local coefficient system $M$, or, what amounts
to the same thing, a $\pi_1(K)$-module~$M$ [AH].  This is the same
thing as ordinary constant 
coefficients on the universal cover~$\tilde K$.
To define color on the other generations of quarks, one 
pulls back the coefficient system to $\tilde K$ via the covering
map $\tilde K\to K$.}

  This requires
a choice of base point, which we take to be a lift $P_0$ of the bottom
vertex.  This determines the other vertices $P_1, P_2$ and the lifts
of the edges $a_i, b_i$, $i=0,1,2$, where an edge is labelled by its
initial vertex.  The situation is shown below in Fig.~3.  
\footnote{We calculate the chain 
in $C_2(\tilde K)$ associated to the lift of the disc diagram shown in Fig.~3
as follows.  In the free group on edges of $\tilde K$ we have, reading around
the diagram once counterclockwise from the base point~$P_0$,
$(a_0a_1b_2)(\bar b_2\bar b_1 \bar a_0)(a_0b_1a_2)$.
The first term may be identified with a lift $\tilde u$ of the up quark
and we shall identify the second with a lift $\tilde d$ of the down
quark.  For the third term, $a_0b_1a_2=\bar a_2(a_2a_0b_1)a_2$,
so as conjugacy classes in the free group
we have
$[a_0b_1a_2]=g^2[a_0a_1b_2]$, where $g$ is the generator of $G=\Z/3\Z$
given by $a\to [1], b\to [1]$.
Thus the 2-cell corresponding to 
$[ a_0b_1a_2]$ maps onto the up quark, but this is
different from the
conjugacy class $[a_0a_1b_2]$;
so the 2-cells are different.  In $C_2(\tilde K)$ we have
the 2-chain determined by the lift of the diagram of Fig.~3 is
$\tilde u+\tilde d +g^2\tilde u$.  
It would appear that there is a mixing of generations of the up quark
in the lift.
If we take into account the face at infinity of Fig.~3, then we
obtain the 2-cycle on $\tilde K$ given
by $\tilde u+\tilde d +g^2\tilde u+\tilde e$, where
$\tilde e$ is a lift of the electron.
}

\vskip.1in
\epsfxsize=3truein
\centerline{\epsfbox{fig3.epsi}}

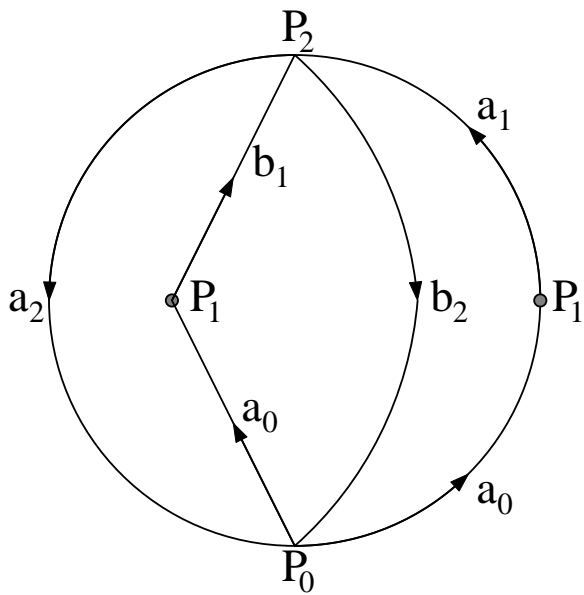
\centerline{Figure 3}

The rules for coloring a diagram are the following, where a coloring is
an assignment of elements of $M$ to the faces, so that the face with
the opposite orientation is assigned the negative of the color of that
face.

\ni 1. A lepton is always white, so has $0$ for color.

\ni 2. The color of a quark is always one of $r,g,b$ and the
color of an antiquark is the negative of the color of the corresponding
quark.  

\ni 3. For a particle to be observable, the sum of the colorings 
of the faces must be 0.

Thus one possible coloring of the diagram~Fig.~3 is to assign the faces
from left to right the colors $r$, $g$, and $b$, whose sum is $0$
(they lift to different faces in $\tilde K$), and assign colors arbitrarily
to the other 2-cells of $\tilde K$, subject to rule~1.
Call such a coloring $f$.
Note that this van Kampen diagram in $\tilde K$ can be viewed as a
spherical diagram where the fourth face is at infinity and corresponds
to the lift of an electron.  As such the coloring extends to one of the
spherical diagram where again the sum of the colors assigned to the faces
is $0$.

So we can reformulate the definition of a coloring of the diagram $D$
for a fermion
in $\tilde K$ as the pull-back to $D$ of a 2-cochain $f\in C_2(\tilde K,M)$ 
with values in~$M$,
 where all leptons
are assigned $0$, where all lifts of quarks in the diagram 
are assigned colors from
$r,g,b$, and where the sum of the values assigned to the faces of
$D$ is $0$.  The rules for a spherical diagram are the same.  So
the cochain $f$ in the previous paragraph can be considered by pull-back 
as either
defined on the disc or as a cochain $\hat f$
on the sphere containing the disc obtained, in this example, 
by extension by~$0$.

The main point here is how to change colorings, that is, to change
cochains in such a way as to preserve rules 1--3 and in a topologically
meaningful way.  This is done by means of coboundaries of 1-cochains.
For example, in the coloring~$f$ indicated above, we can interchange
the colors of the first two faces by $\d h$, where $h$ is the 1-cochain
which is zero on all edges except on $b_1$, and where $h(b_1)=-r+g$.
To interchange the colors of the second and third faces, we
use $\d k$, where $k$ is zero on all edges except on $b_2$, where
$k(b_2)=g-b$.
That rule~3 is preserved is checked directly or, more fundamentally,
follows from the fact that $<\d c, z>=0$ for all 1-cochains $c$ 
and all 2-cycles $z$ (applied to the spherical diagram).
In this way all colorings consistent with the rules are obtained.


These 1-cochains correspond physically to gluons and are the source of the
strong nuclear force.  We shall offer another interpretation of gluons
in a subsequent article on bosons, more in keeping with the ideas of
\S 3 of processes applied to a fermion.

\heading{\bf \S 5. Standard Model}
\endheading

In this section we propose our version of the standard model~SM.
There are alternative versions of SM, some of which were proposed in order to cope with the possibility that the Higgs
boson might not exist.  Now that the data confirm its existence
with better than 99\% probability, we shall eschew treating these
alternative theories from our point of view.

 We define a
{\it fundamental domain\/}
to be one of the two spherical
diagrams shown in Figures~1 and~2 or one obtained from it
by replacements $a,b \to b,a$, $a,b \to \bar a, \bar  b$, or
$a,b\to  \bar b,\bar a$.  Here replacing
$a,b$ by $\bar a,\bar b$, etc., in a diagram means to reverse the
arrows.  
We also admit as fundamental domains completely reducible
\footnote{A diagram is called {\it reducible\/} if it contains a pair of
faces with an edge in common so that the faces are mapped mirror-wise
across that edge.  The faces themselves are called a reducible pair.
In a reducible diagram one can remove the reducible pair and the edge
between them and sew up to boundary of the hole created 
to obtain a new diagram with
two fewer faces.  A diagram is called {\it completely
reducible\/} if its faces can be paired off into reducible pairs.
}
two-faced and four-faced spherical diagrams consisting of
cells $e_\k$ and their oppositely oriented cells $-e_\k=e_{\bar\k}$
(these degenerate\footnote{These degenerate diagrams play an important role
in the sequel article on bosons.} 
examples have corresponding 2-chains~0
in $C_2(K)$).  So in a four-faced
completely reducible fundamental domain one can order the four 2-cells
$\a,\beta,\g,\d$ so that $\a$ and $\beta$
form a reducible pair
 across a common edge
, and $\g$ and $\d$
form a reducible pair across a common edge.
\footnote{It is also possible for both $\a$ and $\beta$ to be a reducible
pair and for $\g$ and $\d$ to be reducible, all at the same time,
as happens in the case of the hypothetical graviton, 
to be considered in the sequel
article
on bosons.  In Appendix~5 below I consider 
completely reducible 6-faced diagrams in order to treat the
$\Delta^{++}$ baryon, as a result of a challenge
from Professor Wu.  Presumably this could continue with
completely reducible $2n$-faced diagrams in order to handle
higher resonances.
}

It should be remarked that not all 2-faced spherical diagrams are considered fundamental domains.  For example, one can take
the electron neutrino $e_\k$ with $\k=[bbb]$, rotate it
a third of a revolution, and glue the two together.  This is
a perfectly acceptable spherical diagram, and indeed it plays
a key role in \S 6 below.  However it is not considered a
fundamental domain because it is not reducible.  The rotation
prevents any edge from acting as a mirror.

A {\it fundamental process\/} for particle physics is
the lift $\tilde f: S\to \tilde K$ of a spherical diagram
$f:S\to K$, where $S$ is a fundamental domain.  A 
{\it fundamental
particle\/} is a triple $(D, \tilde f, h)$, where $\tilde f:S\to \tilde K$ is
a fundamental process lifting $f:S\to K$, where $D\subset S$
is a van Kampen diagram, and where $h\in C^2(\tilde K, M)$
is a coloring in the color group $M$ satisfying

\ni 1. the pull back $\tilde f^*(h)\in C^2(S,M) $ vanishes on
leptons, and

\ni 2. the values of $\tilde f^*(h)$ on quarks are in $\{r,g,b\}$
and the values on antiquarks are in $\{-r, -g, -b\}$.

\ni The fundamental particle is {\it observable\/} if in addition

\ni 3. the sum of the values of $\tilde f^*(h))$ on the faces 
of $S$ is $0$; the invariant formulation
\footnote{The notation $<x,y>$ means to result of evaluating
a cochain~$y$ on the chain~$x$.} of this is that
$<[S],\tilde f^*(h))>=0$, where $[S]$ here denotes the fundamental
2-cycle of the sphere
\footnote{$[S]\in C_2(S,\Z)$ is the sum of the oriented 2-cells in cell decomposition
of the 2-sphere~$S$.  It is readily checked that $[S]$ is a 2-cycle.}

Furthermore, we demand that
any diagram $D'$ obtained from $D$ in 
the triple $(D, \tilde f, h)$ by diamond moves be observable
if the original triple is observable (the 2-cells of $D'$ are mapped
in the same way as those of $D$, so condition~3 is preserved).

Another way of formulating the last condition is to give
a van Kampen diagram $\tilde f:D'\to \tilde K$ and 2-cochain
$h\in C^2(\tilde K,M)$ such that after a sequence of diamond
moves on $D'$ one obtains either a 
fundamental particle~$D$ or  a spherical diagram $S\to \tilde K$
satisfing 1--3.  In the latter case, if $D=S$, then the boundary
label of $D'$ must have been freely trivial, so that $S$ was
obtained by sewing up the boundary completely.  
In all cases,
$D'$ has at most four 2-cells, where the case of four 2-cells
is the case of the freely trivial boundary label (this is also
the case where the particle is its own antiparticle).  


Note that the invariant formulation of condition~3 above is automatically
satisfied for a coboundary~$h=\delta c$, with $c\in C^1(\tilde K,M)$;
that is, $<\tilde f_*[S],\delta c>=0$, because
$[S]$ is a cycle and 
$<[S],\tilde f^*(h)>=<\tilde f_*[S],\delta c>=<\partial \tilde f_*[S],c>
=<\tilde f_*(\partial [S]),c>=0$ by adjointness.  
The main result of this section is a converse.

\proclaim{Theorem} If we let $k=\tilde f^*(h)$, with $\tilde f:S\to \tilde K$ and 
$h\in C^2(\tilde K, M)$ as above,
and if 1--3 are satisfied, then $k=\delta \tilde f^*(c)$ for some cochain 
$c\in C^1(\tilde K,M)$.
\endproclaim

\demo{Sketch of proof}  
We sketch the argument for the proton, namely, Figure~3.  The cases of
the neutron and the antiparticles follow by applying the symmetries 
of interchanging $a$ and~$b$ and of reversing all arrows, as in \S 3.

Since we have to consider spherical diagrams, we consider Figure~3 together
with the face at infinity, which is a lepton.  We pick a coloring of the
finite faces, say $x,y,z$ where these are colors chosen without
repetition from $[0],[1]$ and~$[2]$, read from left to right.  The color
of the face at infinity is white, namely~$0$.  Then the conditions on
$c$ restricted to $\tilde f(S)$, the image of $S$ under $\tilde f$, are 
four linear equations in five unknowns $c(a_0), c(a_1),c(a_2),c(b_1),c(b_2)$
on the left side
(referring to the notation of Figure~3), with variables $x,y,z$ on the
right side.  For example, the first of these equations, corresponding to
the face at infinity, is $c(a_0)+c(a_1)+c(a_2)=0$.
When we row reduce the system, the last row of the row reduction
corresponds to the equation 
equation $0=x+y+z$.  But this compatibility condition is satisfied because
$x+y+z=[0]+[1]+[2]=0$, by the definition of $M$.  So the equations
are solvable.

This determines $c$ on $\tilde f(S)\sk 1$.  Then we extend $c$ in any way to
all of $\tilde K\sk 1$, thereby completing the argument.
\enddemo


\demo{Remark}  Electric charge of a composite particle is given by the coboundary of a 1-cochain~$h$ on $K$.  Namely define $h$ to be
$\frac 1 3$ on an edge labeled $a$, $-\frac 1 3$ on the oppositely
oriented edge, and $h$ of an edge labeled $b$ with any orientation
is defined to be~$0$.  Then $\delta h\in C^2(K,\R)$ calculates the
net electric charge of any van Kampen diagram.  In observable
particles the electric charge must be an integer.  Note that to calculate color, we must use cochains on $\tilde K$
(the two faces in Figure~2 that represent up quarks have different
colors in the proton because they lift to different 2-cells in $\tilde K$), whereas
electric charge can be defined from the projection into $K$.
\enddemo

\demo{Remark}  The weak~isospin~$T_3$ of particle in SM
\footnote{
http://en.wikipedia.org/wiki/Weak{\_}interaction}
governs how it interacts in the weak interaction.
It is analogous to electric charge and is given by
the coboundary of the 1-cochain~$w$ determined by
$w(a)=\frac 1 6$, $w(\bar a)=-\frac 1 6$,
$w(b)=\frac 1 6$, $w(\bar b)=-\frac 1 6$.
In the higher generations, one projects first from
$\tilde K$ to $K$ and then applies $w$ to determine
the weak isospin.

Consequently, three of the fundamental forces of
nature, {\bf the strong and weak nuclear forces and
the electric charge, are determined by coboundaries
of 1-cochains\/} with different value groups.  This is
a remarkable unifying property and serves to 
strengthen our case that {\bf SM is the 
algebraic
topology of the covering map $\tilde K\to K$}.
\footnote{See the Appendix below for a historical
perspective on this assertion.}
There is no evidence at present that gravity fits
into this scheme, and indeed it would be most
remarkable (and stretch the imagination nearly
to breaking point) if
it too were determined by the coboundary of a 1-cochain.
\enddemo

As a last tidbit for this section in response to requests
we draw an avatar of Higgs boson below in Fig.~4.  The argument
that this is correct is deferred to the sequel article on bosons.


\vskip.1in
\epsfxsize=3truein
\centerline{\epsfbox{fig4.epsi}}

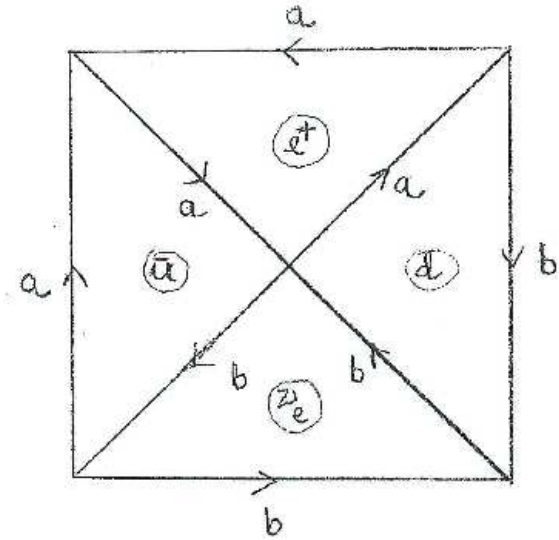
\centerline{Figure 4}

\heading{\bf \S 6.  Neutrino oscillations}
\endheading

This section gives a physical interpretation for the processes
in $Z_2(\tilde K)=\pi_2(K)$ which are not lifts of processes from
$Z_2(K)$.  
  For many years it was a mystery that the observed neutrinos
coming from the sun were a third of the predicted value.  The
explanation finally provided was that a neutrino in free space
oscillates between the three types, electron neutrino~$\nu_e$,
muon neutrino~$\nu_\mu$, and tau neutrino~$\nu_\tau$.
Since only $\nu_e$ was observed in the experiment, what we
observe is only a third of the actual production from nuclear
reactions (due to $\beta$-decay) in the sun.

We begin by noting that, by an Euler characteristic computation,
the rank of $Z_2(\tilde K)$ is 8.
We can account for 6 free generators by lifts of the classes
in $\pi_2(K)$ represented by the spherical diagrams in
Figures~1 and ~2.  We shall now account for the other two
free generators.  

Let $\tilde \nu$ denote a lift of the 2-cell $e_\k$ in $K$,
where $\k=[bbb]$.  Let $g$ be a generator for the covering
group $G\cong \Z/3\Z$ of $\tilde K$.  Then an easy calculation
shows that
 $g\tilde\nu-\tilde \nu \in Z_2(\tilde K)=H_2(\tilde K)$.  This 
class is in the kernel of the projection $H_2(\tilde K)\to
H_2(K)$, so it is undetectable in $K$.  We also have
the class $ g (g \tilde \nu - \tilde \nu)=g^2 \tilde \nu - g\tilde \nu$
in the kernel.  A calculation shows that these two chains
are linearly independent in $C_2(\tilde K)$ and hence in
$Z_2(\tilde K)$.  It follows that we have accounted for the
extra two free generators of $Z_2(\tilde K)$ that do not
arise from lifts of cycles in $K$.
\footnote{$Z_2(\tilde K)$ maps onto $Z_2(K)$, as follows
from the Proposition of \S 3, and $Z_2(K)$ is free abelian,
so the kernel is a direct summand; it follows that it is of
rank~2.
}

Now recall our premise from \S 3 that processes involving
fermions arise from 2-cycles.  Thus $g \tilde\nu- \tilde \nu$,
 $g^2 \tilde \nu-g\tilde \nu$, 
 and $\tilde \nu-g^2\tilde \nu$ are processes; these
processes exchange the three neutrinos $\nu_e, \nu_\mu$
and $\nu_\tau$ in cyclic fashion, accounting for the
oscillation of neutrinos.  Of course we cannot deduce from
the topology alone the rates of the processes, so this is
all the information we can hope to gain from this 
type of argument about neutrino oscillation.

Now we can replace $b$ by $a$ in the argument above
and deduce that there exist processes which exchange
cyclically the electron, muon, and tau particle.  Such
oscillations
have not yet been detected in the laboratory.

There is fortunately a check for the correctness
of these ideas, as follows.
The process by which the muon decays is 
\footnote{See http://en.wikipedia.org/wiki/Muon
}
$\mu^-\to \nu_\mu+e^-+\bar \nu_e$.
Now $\mu^-=g\tilde e$, where $\tilde e$ is the lift of the electron
to $\tilde K$ and $g$ is as above a generator for the covering group.
Also $\nu_\mu=g\tilde \nu_e$, where $\tilde \nu_e$ is the lift of $\nu_e$,
the electron neutrino.  Bringing all terms to the left side,
the decay process above is compatible with our theory if
$g\tilde e-g\tilde\nu_e-\tilde e+\tilde \nu_e$ is a 2-cycle in $Z_2(\tilde K)$.
But this last expression is $(g-1)\tilde e-(g-1)\tilde\nu_e$, which
is indeed a 2-cycle.

\heading{\bf \S 7.  Questions and comments}
\endheading

\ni 1. Shupe's "no mixing rule" ([Sh] p. 88), which states, on our terminology,
that $aa\bar b$,$a\bar b\bar b$, and their
inverses do not occur in the construction of the complex $K$, is 
at the heart of this paper.  In the sequel paper on bosons, we 
offer a justification for this rule.

\ni 2.  There is an uncanny symmetry in this model.  Namely the Klein
four-group, which appears as the subgroup of $\Aut(F)$ consisting
of the identity, $\{a,b\}\to\{b,a\}$, $\{a,b\}\to\{\bar a,\bar b\}$,
and $\{a,b\}\to \{\bar b,\bar a\}$, acts on the theory. 
 This symmetry demands further explanation.

 \ni 3.  The definition of the 2-complex~$K$ looks arbitrary although
it arose naturally from a reinterpretation of the work of [Ha] and [Sh].
I can prove the following result which may serve to motivate it
for mathematicians.

\ni{\bf Theorem.}  There exists an infinite dimensional CW-complex~$L$
with finite $n$-skeleton for all $n\ge 0$ with the following properties.
\roster
\item $L$ is contractible; in particular it is simply connected.
\item $L$ is acted on freely and cellularly by the group~$G=\Z/3\Z$;
hence the projection $L\to L/G$ is the universal covering.
\item $L\sk 2=\tilde K$.
\item $L\sk{2n+1}$ is homotopy equivalent to $S^{2n+1}$ for all $n\ge 1$.
\endroster

\ni The argument makes use of the periodic resolution for
$\Z$ over $\Z[G]$, which is modified in low dimensions 
($\le 3$) to 
account for the unusual presentation $P$ for $G$.  

It follows from the theorem that $L$ is homotopy equivalent
to $BG$, the classifying space of principal $G=\Z/3\Z$ fibrations,
which is
a space of type~$K(G,1)$.

\ni{\bf Question.} Can $\tilde K$ be equivariantly imbedded
in $S^5$ for the $G=\Z/3\Z$-actions?  Here $G$ acts on $S^5$
by the diagonal action of third roots of unity on the unit vectors
of $\C^3$.  In the sequel paper we show that $\tilde K$ imbeds
equivariantly in $S^{2n+1}$ for $n$ sufficiently large ($n\ge 6$).

 \heading{\bf Appendix 1.  Historical remark}
 \endheading

 C-N Yang [Ya] relates a conversation he had with Andre Weil in which
 the latter suggested that the particles that were being discovered
 might be explained by ``geometry and topology".  Yang writes that he did not
 understand this, and no one at the meeting thought
 to ask him what he meant. As far as Weil's suggestion about geometry, that
 has proved to be correct in the development of Yang-Mills theory.
 But there is no precedent in the literature that what he had in mind
 may also have included pure topology.

 \heading{\bf Appendix 2.  Color group $M$ and $A_2$}
 \endheading

 This Appendix will relate the color group $M$ to $A_2$,
 which is the root system for SU(3); this lie group is known
 to physicists as the theory that describes the strong
 nuclear force. 
 
  The notation follows \S 4 for the definition
 of $M$.  As usual $G$ denotes the cyclic group of order~3
 and $g$ is a generator.  We make the complex numbers~$\Bbb C$
 into a $G$-module by letting $g$ act by multiplication by
 $\z=e^{2\pi i/3}$, a primitive third root of unity.  Geometrically
 this represents a rotation through 120 degrees.
 
 \proclaim{Theorem} The map $g\to \z$ extends to an
 homomorphism of $G$-modules $\phi:\Z[G]\to \Bbb C$
 with the properties that
 \roster
 \item the image of $\phi$ is $\Z[\z]$, the subring of $\Bbb C$
 generated by $\z$ and the unit element,
 \item the kernel of $\phi$ is $N\Z[G]=(1+g+g^2)\Z[G]$, so
 \item $\phi$ induces an isomorphism $M\to \Z[\zeta]$, and
 \item the images of the colors $\pm [j]$, where $j=0,1,2$,
 form the root system $A_2$; thus $\phi( \pm[j])=\pm\z^j$
 comprise the set of sixth roots of unity in $\C$.
 \endroster
 \endproclaim

  For example let us prove
 that 
the kernel of $\phi$ is the ideal generated by $N=
 1+g+g^2$ in $\Z[G]$ and 
 that $\phi$ induces an isomorphism $M\to \Z[\zeta]$. 
  The element~$N$  is in the kernel
 since $\z$ is a primitive third root of unity, so
 satisfies the polynomial $1+z+z^2=\frac {z^3-1}{z-1}$.
 The rank of $\Z[\z]$ is 2, since it is the set of
 algebraic integers
 in an algebraic number field $\Q[\z]$ of degree~2 over
 the rationals~$\Q$.  So a rank count shows that the induced
 map $M\to \Z[\z]$ is an isomorphism.
 
 \def\Hom{\hbox{\rm Hom}}

\def\Hom{\hbox{\rm Hom}}

\heading{\bf Appendix 3. Conservation laws in SM and quantum numbers}
\endheading

In \S 1 it was pointed out that the fundamental conservation
law involving preons is that the net numbers of $a's$ and $b's$
are individually conserved in a reaction predicted by SM.  From
this followed the result that there are only three generations
of fundamental fermions of SM, {\it etc.\/}

In this appendix we shall define conservation laws and quantum
numbers for elementary particles and show that, if we consider
lifts to the universal cover~$\tilde K$, then the preon number
for a lift~$a_i$ or $b_i$ of a preon is a quantum number.  It follows that
 there is no theoretical reason that one cannot
detect the preon number directly. 

\demo{Definition}  A {\it conservation law\/} in $A$, where
$A$ is a ring with unit, is a 
2-cochain~$c\in C^2(\tilde K,A)=\Hom(C_2(\tilde K,\Z),A)$
such that $<z,c>=0$ for all $z\in Z_2(\tilde K,\Z)$.  Here
$<z,c>$ is the result of evaluating $c$ on the chain $z$ 
to yield an element
of $A$.  A {\it quantum number\/} for the conservation
law~$c$ is the result of evaluating $c$ on a particle
(that is, on a van Kampen diagram in $\tilde K$).
\enddemo

The reason this is called a conservation law is that a 
reaction is given
 by $\k_1+\k_2+\dots+\k_p\to\lambda_1+\dots+\lambda_q$,
so it follows that $c(\k_1)+c(\k_2)+\dots+c(\k_p)=c(\lambda_1)+
\dots+c(\lambda_q)$.
\footnote{$\k_1+\k_2+\dots+\k_p-\lambda_1-\dots-\lambda_q$
is a 2-cycle on $\tilde K$, whence $c$ vanishes on it.
}
By {\it adjointness\/}, it follows that, if 
$c=\delta h$ with $h\in C^1(\tilde K,A)$,
then $c$ is a conservation law. 
\footnote{If $z$ is a 2-cycle on $\tilde K$, then
$<z,c>=<z,\d h>=<\partial z,h>=0$.
}
 Thus it follows from the
remarks of \S 5 that electric charge and weak isospin~$T_3$ are quantum numbers with
 values in $\R.$
 The next result says the converse is true.

\proclaim{Proposition}  If $c\in C^2(\tilde K,A)$ is a
conservation law,
 then $c=\delta h$ for some $h\in C^1(\tilde K,A)$.
 Thus the set of conservation laws in $A$ is $B^2(\tilde K,A)$.
\endproclaim

\demo{Proof}  
We are given $c\in C^2(\tilde K,A)=\Hom(C_2(\tilde K,\Z),A)$ such that $c$
vanishes on $Z_2(\tilde K,\Z)\subset C_2(\tilde K,\Z)$.

By the universal coefficient theorem
\footnote{http://en.wikipedia.org/wiki/Universal$\_{\null}$coefficient$\_{\null}$theorem},
$H^2(\tilde K,A)=\Hom(Z_2(\tilde K,\Z),A)$,
where we have used the facts that $H_1$ vanishes (since $\tilde K$
is simply connected) and $H_2(\tilde K,\Z)=Z_2(\tilde K,\Z)$
since there are no 3-cells.
Also $Z^2(\tilde K,A)=C^2(\tilde K,A)$ since every 2-cochain is
a 2-cocycle.
Thus
$H^2(\tilde K,A)=C^2(\tilde K,A)/B^2(\tilde K,A)=\Hom(Z_2(\tilde K,Z),A)$.
This may be expressed as the short exact sequence
$$0\to B^2(\tilde K,A)\to C^2(\tilde K,A)\to \Hom(Z_2(\tilde K,Z),A)\to 0.$$
Also we have the surjective map $C^1(\tilde K,A)\epi B^2(\tilde K,A)$,
which may be spliced with the first map in the short exact sequence to
yield the exact sequence
$$
C^1(\tilde K,A)@>\delta >>C^2(\tilde K,A)\to 
\Hom(Z_2(\tilde K,Z),A)\to 0.$$

Now $c\in C^2(\tilde K,A)=\Hom(C_2(\tilde K,\Z),A)$
and the map on the right in the last exact 
sequence is obtained by restricting a homomorphism
to $Z_2(\tilde K,\Z)$ and then considering the image 
to lie in $A$ via the map $\Z\to A$.
It follows that the image of $c$ in $\Hom(Z_2(\tilde K,\Z),A)$
is zero, and, by exactness, $c=\delta h$ for some $h\in
C^1(\tilde K,A)$.\qed
\enddemo

Let us specialize now to the case $A=\R$.  
Then $C^2(\tilde K,\R)$ is of dimension~12
while the rank of $Z_2(\tilde K,\Z)$ was calculated
in \S 6 to be 8.  It follows that the dimension
of $B^2(\tilde K,\R)$ is 4, and hence there are
4 linearly independent conservation
laws in SM.
\footnote{The same result holds for every field.}

Let us focus attention on one that has not been
mentioned before.  Let $f\in C^1(\tilde K,\R)$
be given by $f(a_i)=\frac 1 3$, $f(b_i)=-\frac 1 3$,
$f(\bar a_i)=-\frac 1 3$, $f(\bar b_i)=\frac 1 3$,
for $i$$\pmod 3$.
If $h$ and $w$ are 1-cochains on $\tilde K$ determining
charge and weak isospin $T_3$ respectively
(in the notation of \S 5),
then $f=2(h-w)$.  This identifies $\delta f$ with the
weak hypercharge $Y_W$.
\footnote{http://en.wikipedia.org/wiki/Hypercharge.
In the form $Q=I_3+\frac 1 2 Y$ the relation is attributed to
Gell-Mann and Nishijima.}

We can calculate the vector space $B^2(\tilde K,\R)$ 
of conservation laws over $\R$
as follows.  We have $B^2=C^1/Z^1$ where $Z^1=B^1$.
From these relations we see that $Z^1(\tilde K,\R)$ is the
set of functions~$f$ on the 1-cells $a_i,b_i$ of $\tilde K$
so that $f(a_i)=f(b_i)$, $f(a_0)+f(a_1)+f(a_2)=0$, for $i\pmod 3$.
This is a 2-dimensional subspace of the 6-dimensional
space~$C^1$,
so the 4-dimensional quotient $B^2$ can be effectively determined.
Here is the result of this calculation.

If we let $x$ be one of $a_i, b_i$, $i\pmod 3$, we define 
$\D_x\in C^1(\tilde K,\R)$ by 
$\D_x(x)=1, \D_x(\bar x)=-1$, and $\D_x(y)=0$ for all
other $y$.  Then for each lift $\tilde \k$ of an elementary fermion 
$\k$
\footnote{So $\k$ is either an electron, neutrino, quark or an 
antiparticle of one of these.
}
 we have $\d\D_x(\tilde \k)= +1$ if the boundary label
of $\tilde \k$ contains $x$, $\D_x(\tilde \k)=-1$ if the boundary
label contains $\bar x$, and $\D_x(\tilde \k)=0$ otherwise.

\proclaim{Proposition}  For each $x\in\{a_i,b_i\}$, $i\pmod 3$,
 $\d\D_x$ is a conservation law,
and for each lift $\tilde \k$ of an elementary fermion
$\d\D_x(\tilde \k)$ is a quantum number.
For any collection of 4 of the 6 lifts $a_i,b_i$ the corresponding
conservation laws are linearly independent.\qed
\endproclaim

\demo{Remark}  The proposition provides a test for the
structure of the proton~$p$ proposed in Figure~2.  
For any lift~$\tilde p$ of $p$ to $\tilde K$, we see
that $\d\D_{a_i}(\tilde p)=1$ for all $i\pmod 3$.
This structure also has the consequence that the two lifts
of $u$ in $\tilde p$ must be of different generations,
as was pointed out in the footnote of \S 4.
If we tried the structure $z=[a_0a_1b_2]+
[\bar b_2\bar b_1\bar a_0]+[a_0a_1b_2]$ instead
(which projects to $2u+d$ but is not the continuous
lift given by covering space theory), we would
find $\d\D_{a_0}(z)=1, \d\D_{a_1}(z)=2,
\d\D_{a_2}(z)=0$.
So these new quantum numbers may be useful
tools in determining the structure of composite
particles as  van Kampen diagrams.

\enddemo

\heading{\bf Appendix 4.  $\beta$-decay revisited}
\endheading

When I explained to a physicist how van Kampen diagrams
 gave a consistent interpretation
to the ideas of [Ha] and [Sh], which solved many of the open problems these
papers
posed, he expressed interest in seeing in detail how $\beta$-decay works
from the topological point of view (the point of view of Feynman diagrams
is well-known 
\footnote{http://en.wikipedia.org/wiki/Beta{\_}decay}
$d\to u+e^-+\bar\nu_e$
but that leaves unexplained what is actually happening at the vertices
of the Feynman diagram).

Examine Figure~5 below.

\vskip.1in
\epsfxsize=3truein
\centerline{\epsfbox{fig5.epsi}}

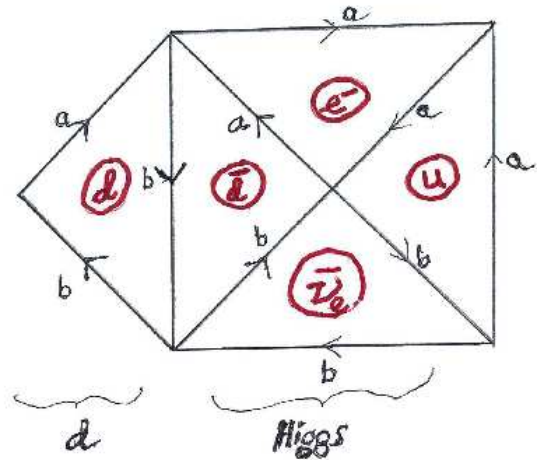
\centerline{Figure 5}

\vskip.1in

The left half of the diagram is the down quark~$d$ whereas the right
half is the Higgs boson.  The latter is obtained from the spherical
diagram Figure~1 by making two cuts along edges labeled
$a$ and $b$ and splaying the result out
as a van Kampen diagram in the plane; so this yields is a 2-cycle and
acts on the van Kampen diagram for $d$ in the manner described in
\S 3 following Remark~1.
\footnote{This description is purely topological and ignores entirely
the question of magnitude of energies involved.  So the Higgs in question
is of the nature of a virtual particle, arising from the vacuum state only
ephemerally and disappearing immediately after interacting with the 
down quark.}

The faces labeled $d$ and $\bar {d}$ are mapped mirror-wise across their
common edge, so we may do a reduction, removing the pair and that edge,
and sew up the hole to create the diagram Figure~6 below.

\vskip.1in
\epsfxsize=3truein

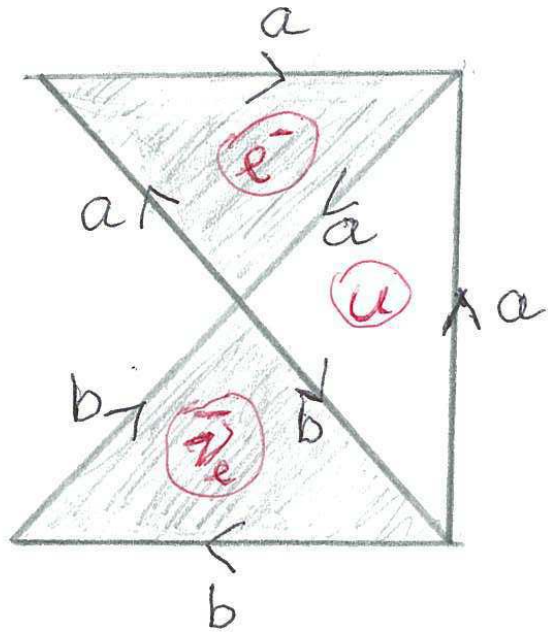
\centerline{Figure 6}
\centerline{\epsfbox[225 485 425 715]{fig6.epsi}}

The two shaded faces are $e$ and $\bar \nu_e$ which are now shed
to leave the up quark~$u$.  
So the net result is the
reaction $d\to u+e^-+\bar\nu_e$.  As a parenthetical remark, we observe that
the union of
those two shaded faces constitute the $W^-$ boson,
 so the reaction can be rewritten as the composite of two reactions,
$d\to W^-+u$ and $W^-\to e^-+\bar\nu_e$.

\heading{\bf Appendix 5.  $\Delta^{++}$ baryon}
\endheading

 Professor Wu challenged me to describe
the $\Delta^{++}$ baryon
\footnote{
en.wikipedia.org/wiki/Delta{\_}baryon
}
in my terms.
The result is shown below in Figure~7.

\vskip1in
\epsfxsize=3truein
\centerline{\epsfbox[60 470 535 715]{fig7.epsi}}

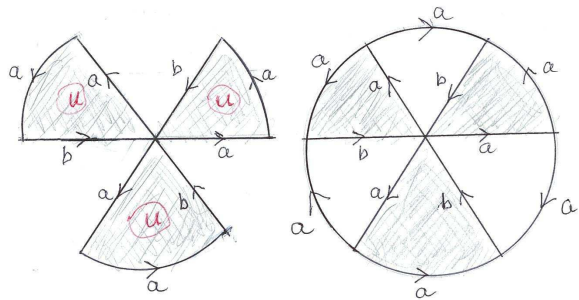
\centerline{Figure 7}

 It shows three
$u$ quarks with a single vertex in common (like the radiation
hazard sign).  The spin of the resulting particle is $\frac 3 2$,
while the remarks of \S 5 show how to calculate that the electric
charge is~2 and the weak isospin~$T_3$ is $\frac 3 2$.
The second diagram on the right shows how to imbed $\Delta^{++}$
in a diagram with 6 faces and freely trivial boundary label, so
the enlarged diagram folds up to a spherical diagram and
represents a 2-cycle.  That 2-cycle is 0, for the diagram is
completely reducible.

All other spherical 
diagrams previously considered here have 2 or 4 faces,
so this is the first where 6 faces are needed.
That fact by itself makes Professor Wu's question
interesting and motivated me to include it  here.

\heading{\bf Appendix 6. Quantum numbers and the first Chern class}
\endheading

In \S 5 electric charge was interpreted as the coboundary of
a 1-cochain on $\tilde K$ which is evaluated on particles
(that is, on van Kampen diagrams).  This is satisfactory from
the point of view of algebraic topology but it is
desirable to have a direct geometric interpretation.
 The answer given below is that charge, 
up to a factor of $\frac 1 3$, is
the first Chern class of an explicitly constructed
line bundle.
This involves an interpretation of the preons $a$ and $b$
as sections of a complex line bundle,
which is then pursued to deal with other quantum numbers.

Let $D$ be the van Kampen diagram for a fundamental fermion of
SM, so $D$ represents a positron, an
up quark, an anti-down quark, or 
an electron neutrino,.  We view $D$ as a map to $K$; as
pointed out in \S 5, charge is defined on $K$ and pulls back
to $\tilde K$ via the covering map $\tilde K\to K$.  
Thus the boundary label of $D$ is either $aaa, aab, abb,$ or $bbb$.
The cases of their antiparticles involve only a change of 
sign in the construction.  Thus the boundary of $D$ is subdivided into
3 segments, each of which is mapped onto an edge $a$ or $b$ of $K$, and the
end points of the segments are mapped to the unique vertex of $K$.

Consider the trivial $U(1)$-bundle $D\times S^1$ over $D$.  
Here $U(1)=S^1$ is considered as the unit interval $I=[0,1]$ in $\Bbb R$ with
end points identified, $S^1=I/{0\sim 1}$.  Define
a section~$\sigma$ of the bundle 
over the boundary $\partial D$ of $D$ by the identity
map of $S^1$ beginning at the base point $0$ over each segment
 labeled~$a$ and by the constant map $0$ over the segment labeled~$b$.
 Note that if we consider this section as a map $f_\sigma: S^1\to S^1$,
 the degree of the map is $3, 2, 1$ or $0$ in the respective cases
 $aaa, aab, abb,$ or $bbb$.

\proclaim{Theorem 1}  The electric charge of the fermion is
$\frac 1 3$ of the first Chern class in $H^2(S^2,\Z)\cong \Z$ 
of the complex line bundle
over $S^2$ with clutching function~$f_\sigma$.
\footnote{For the notion of clutching function, see 
en.wikipedia.org/wiki/Clutching{\_}construction
or Hatcher's book on vector bundles
http://www.math.cornell.edu/$\sim$hatcher/VBKT/VBpage.html,
p.~22.
}
\endproclaim

\demo{Remark}  Equivalently, the electric charge is the
obstruction to extending $\sigma$ to a never-vanishing
 section
of the trivial line bundle $D\times \Bbb C$ over $D$,
up to a factor of $\frac 1 3$;
here $S^1$ is considered as the image of the
exponential map
$t\mapsto e^{2\pi it}$ in $\Bbb C$, $0\le t\le 1$.
\enddemo

\demo{Proof of Theorem}  The first Chern class of the
line bundle is the degree of the clutching function 
$f_\sigma:S^1\to S^1$.
\enddemo

Next we generalize the construction just made
to other quantum numbers. 
We do the construction in $\tilde K$
rather than in $K$, as we did in the case of electric charge,
which is invariant under deck transformations of the covering
map $\tilde K\to K$, since we want to allow the possibility of
quantum numbers that are not invariant under deck transformations.

 If $\a_i$ and $\bb_i$ are 
integers, $i\mod 3$, let $F$ be the section of the trivial $U(1)$
bundle over $\tilde K\sk 1$ defined by maps of $S^1$ to itself of
degree $\a_i$ over edges $a_i$ and $\bb_i$ over edges $b_i$.
The obstruction to extending $F$ over all of $\tilde K$ (which
depends only on the integers $\a_i, \bb_i$ and not on representative
maps) can be calculated as follows.  Let $e_{\k_i}$ be one of the 
2-cells of $\tilde K$.  Then the section $F$ is defined over the
boundary $\partial e_{\k_i}$ of $e_{\k_i}$, which is a covering $\k_i$ of one 
of the fundamental fermions $\k$ (so $\k$ is one of $[aaa],[aab],[abb]$
or $[bbb]$ or their inverses).  Let $S_{\k_i}$ be the 2-sphere which
consists of two identical copies of $e_{\k_i}$ glued along their
common boundary by the identity map, and let $L_{\k_i}$ be the 
complex line
bundle over $S_{\k_i}$ given by the clutching function $F$.
Then the obstruction to extending $F$ over $e_{\k_i}$ is the
first Chern class $c_1(L_{\k_i})\in H^2(S_{\k_i},\Z)\isom \Z$.

Thus the obstruction to extending $F$ over all of $\tilde K$ is
the collection of first Chern classes $c_1(L_{\k_i})\in \Z$.  
This amounts to 24 integers, counting lifts of antiparticles
of the fundamental fermions.
Call this 24-tuple of integers $q_F$.

\proclaim{Theorem 2}  For every conservation law $c\in B^2(\tilde K,\Q)$
there is a section $F$ of the trivial $U(1)$ bundle over 
$\tilde K\sk 1$ and a positive integer~$N$ so that 
$\frac 1 N q_F$ is the collection of quantum numbers associated to $c$.
The collection $q_F$ is given by the first Chern classes 
of complex line bundles $L_{\k_i}$ constructed above.
\endproclaim

For the proof one takes $h\in C^1(\tilde K,\Q)$ so that $\delta h=c$,
where $N$ is a number that clears all denominators.  Then one applies
the construction of the preceding paragraph to $Nh$.

\demo{Example} If $h(a_i)=1$, $h(b_i)=0$ for all $i \pmod 3$ and $N=3$, then
we obtain the electric charge lifted to $\tilde K$.
\enddemo

Another way of stating the result is as follows.  Let $X$ denote the
double of $\tilde K$ along $\tilde K\sk 1$.  So $X$ consists of
two disjoint 
copies of the 2-complex $\tilde K$ identified along their common
subcomplex $\tilde K\sk 1$.  Take two copies of the trivial complex
line bundle over $\tilde K$ and identify them over $\tilde K\sk 1$
by the isomorphism given by $F$.  This produces a line bundle $L$ over $X$.
Then $c_1(L)\in H^2(X,\Z)$ contains all the information in the
collection $q_F$.
Namely, if we take the double of the 2-cell $e_{\k_i}$ lying inside
$X$, this is an imbedded copy of the 2-sphere $S_{\k_i}$.
So we can restrict the bundle $L$ to $S_{\k_i}$ and take its
first Chern class, thereby recovering the quantum number up to
the factor of $N$.
\footnote{The construction of the bundle $L$ from the data $F$
mimics the clutching function construction for bundles over
spheres.  It works because $X\setminus \tilde K\sk 1$ is a
disjoint union of cells each of whose closures in $X$ is an 
imbedded disc.  So the argument for characterizing bundles
over spheres by clutching functions works for 
characterizing bundles over~$X$.}

There is a formal way to incorporate the multiplicative factor~$N$
into the answer.  One considers the group $\text{Pic}(X)\otimes \Q$
where $\text{Pic}(X)$ is the Picard group of 
isomorphism classes of complex line
bundles under the operation of tensor product and $\Q$ is the
rational numbers.  The element $L\otimes \frac 1 N$ is then
an invariant of the conservation law $c$ over $\Q$ and incorporates
all the quantum numbers by the operation of taking the
first Chern class.

\demo{Example}  Consider a lift $\tilde e$ of the electron
to $\tilde K$, so $\tilde e$ is a 2-cell with boundary label
 $[\bar a_2\bar a_1\bar a_0]$.  The 2-sphere $S_{\tilde e}$
 is the double of the closed cell~$\tilde e$ over its boundary.
 If we take the electric charge as the conservation law, then
 the map of the boundary $\partial \tilde e$ to $U(1)$ has
 degree $-3$ and $N=3$.  The element $L\otimes \frac 1 3$
 of $\text{Pic}(X)$ restricted to
 $S_{\tilde e}$ is the complex line bundle over $S_{\tilde e}$
 with clutching function a map of degree~$-1$ of the circle to
 $U(1)$.  This is the tautological line bundle over $\Bbb C P^1$
 whose associated principal bundle with fibre $U(1)=S^1$
 is the Hopf fibration over $S_{\tilde e}$.  In terms of algebraic
 geometry, this is the bundle $O(-1)$ over $\Bbb C P^1$.
 
 If instead of the electron we take a lift of the positron,
 the result is the hyperplane bundle $O(1)$ over $\Bbb C P^1$.
 In more colorful language, the charge of the electron is
 the Hopf bundle and the charge of the positron is
 the hyperplane bundle over $\Bbb C P^1$.
\enddemo

\demo{Remark}  The 2-complex $X$, constructed in this section 
as a formal device for relating conservation laws to complex
line bundles, plays a
fundamental role in the sequel paper on bosons.  It will be shown there
that the Higgs mechanism for generating masses of particles in SM
can be understood in terms of the geometry of $X$ and that the
involution interchanging the two copies of $\tilde K$ and fixing
$\tilde K\sk 1$ can be
understood as changing handedness of particles (left for right
and right for left).

As pointed out in the lead paragraph of this section, we have given
a direct geometric interpretation for the preons $a,b$  
as maps of one circle to another (from a closed 1-cell of $K$
to $U(1)$), of degree~1
for $a$ and degree~0 for $b$.  Such an interpretation is lacking
in [Ha] and [Sh], where the preons are merely labels.

An open problem is to explain why nature chose this strange mechanism
$K,\tilde K, X$ for particle physics and why the number~3 as in
3-fold covering space. 

\enddemo

\heading{\bf Appendix 7.  Colorings as conservation laws}
\endheading

In this appendix we give examples of conservation
laws where the coefficient ring~$A$
is the color group~$M$.  We shall use the isomorphic ring
$\Z[\zeta]$ (see Appendix~3) where $\zeta=e^{2\pi i/3}$.

Recall that in this model 
the color set~$A_2$ is the subset $\pm \zeta^j$ where
$j$ ranges over the integers modulo~3.

\demo{Definition}  A {\it  coloring\/} of $\tilde K$
is an element $c\in B^2(\tilde K,\Z[\zeta])$ such that \roster
\item $c$ vanishes on every lepton, and
\item $c$ assumes all the elements of $\{1,\zeta,\zeta^2\}$
as values on the set of lifts
of the up quark~$aab$ 
and also on all the lifts of the down quark~$\bar a\bar b\bar b$.
\endroster
Condition~(1) is means physically that leptons do not
feel the strong force.
 It follows from (2) that $c$ cannot assume the same value
on two different 2-cells of $\tilde K$ which lift the
same quark of $K$.
Since $c$ is a coboundary, $c$ vanishes on all 2-cycles.
 Since $c$ is a cochain, 
$c(\kappa)=-c(\bar \kappa)$, where $\bar \kappa$ is the oppositely
oriented 2-cell to $\kappa$. 
\enddemo

It is not immediately clear that there are any 
colorings (the catch is to show that a candidate cochain vanishes
on all 2-cycles in $Z_2(\tilde K,\Z)$).  That is the content of the next 
result.

\proclaim{Theorem}  
There are six colorings of $\tilde K$, which
are hence conservation laws with values in $\Z[\zeta]$.
\endproclaim


\demo{Proof of theorem}
The crux of the argument consists of constructing a family of
elements of
$B^2(\tilde K,\Z[\zeta])$.  This will be done by explicit 
construction of 2-cochains which vanish on $\pi_2(\tilde K)=
Z_2(\tilde K,\Z)$.
We recall from \S 6 that $Z_2(\tilde K)=:Z_2(\tilde K,\Z)$ is a free
abelian group of rank~8.  Two of the free generators are
leptonic in the sense that they are of the form $g\kappa-\kappa$,
where $\kappa$ is a lift of a lepton and $g\in G\cong\Z/3\Z$, so
any observable cochain will vanish on them.  Thus there are
six remaining linear conditions of vanishing that must be satisfied.

The six vanishing conditions arise from the lifts of
the spherical diagrams represented by Figures~1 and~2.  It is
easier to visualize these in terms of van Kampen diagrams, so
we make cuts along edges to obtain disc diagrams; for convenience
we have shown one lift each of the two disc diagrams below in
Figure~8 (one can check that these are obtained by cutting spherical
diagrams along edges since the boundary labels are freely trivial).

\vskip.1in
\epsfxsize=5truein
\centerline{\epsfbox{fig8.epsi}}

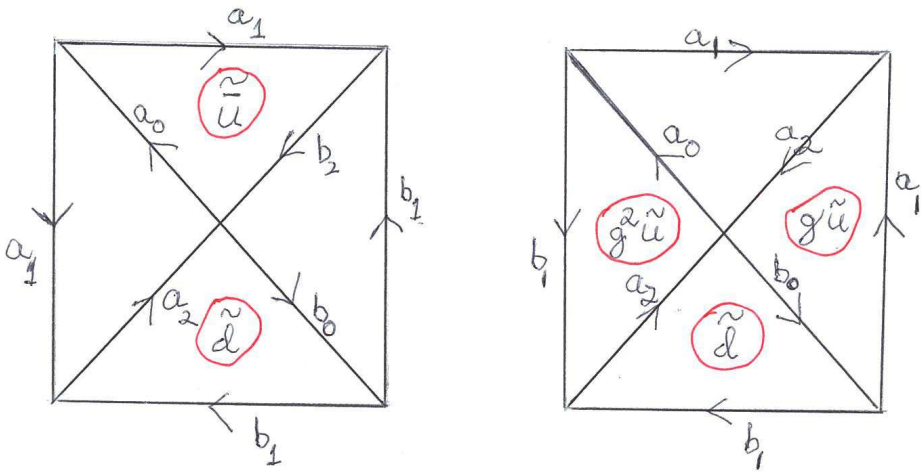
\centerline{Figure 8}

The left figure is a lift of the diagram for the Higgs boson that
participates in $\beta$-decay (see Figure~5 in Appendix~4).  We make
the convention that the bottom triangle labeled~$\tilde d$ represents the
down quark.  It follows that the top triangle labeled $\tilde {\bar u}$
is the anti-particle to the 
up quark (since $\beta$-decay does not mix the generations
of quarks).  The two remaining triangles are leptons.
So the vanishing condition on the cochain~$c$ we are attempting to
construct is $c(\tilde d)=c(\tilde u)$.  Similar conditions must
hold in the other generations, since the process of $\beta$-decay
lifts to these generations (see \S 3).  Thus we obtain
additional conditions $c(g\tilde d)=c(g\tilde u)$ for all $g\in G$.
This gives three of the six linear conditions that $c$ must satisfy to
be an observable cochain.

The three colors $c(g\tilde d)\in \{1,\zeta,\zeta^2\}$, $g\in G$,
 may be chosen
in any manner so as to satisfy that they exhaust this set of colors.
That means there are six choices of potential cochains if we consider
their values restricted to the lifts of the down quark.  But the
conditions $c(g\tilde d)=c(g\tilde u)$ of the previous
paragraph determine completely the cochain~$c$, since the value of
$c$ on leptons is~0.
It remains to establish that each of these six cochains vanishes on
cycles $Z_2(\tilde K)$.

We pass to the diagram on the right in Figure~8.  The bottom triangle
is seen to be $\tilde d$, and a calculation shows that the left
and right triangles are $g^2\tilde u$ and $g\tilde u$ respectively.
\footnote{Recall $g$ is the generator of the 
covering group~$G\cong\Z/3\Z$ where $g$
is given 
by $a\to 1, b\to 1 \pmod 3$.  The action of $g$ on the lifts $a_i,b_i$
is to raise the indices by $1\pmod 3$.}
The top triangle is a lepton, so is assigned 0 by the cochain~$c$
under construction.  
Thus the values of $c$ on the quark 2-cells in the diagram
are $\{c(g^2\tilde u),c(\tilde d),c(g\tilde u)\}=\{c(g^2\tilde d),c(\tilde d)
,c(g\tilde d)\}
=\{1,\zeta,\zeta^2\}$  The same result holds for all translates
of the diagram by $g\in G$.  
But $1+\zeta+\zeta^2=0$ in $\Z[\zeta]$, so it follows that
the three remaining linear conditions are satisfied and that
all six cochains~$c$ vanish on $Z_2(\tilde K)$.
Hence, by the first proposition of Appendix~3, $c\in B^2(\tilde K,\Z[\zeta])$.
 \qed
\enddemo

\demo{Example}  If we consider the coloring~$c$ given by
$d\to 1, gd\to \zeta,g^2d\to \zeta^2,
u\to 1, gu\to \zeta,g^2 u\to \zeta^2$,
then one can check that $c=\delta L$, for 
$ L\in C^1(\tilde K,\Z[\zeta])$,
where $L$ is given by $L(b_i)=0, i\pmod 3$, 
$L(a_0)=1+\zeta^2, L(a_1)=1+\zeta, L(a_2)=\zeta+\zeta^2$.
Other choices of $H'$ with $\delta H'=c$ 
differ from $H$
 by a 1-coboundary.
 
 If we apply a permutation~$\sigma$ of the 
 set~$\{1,\zeta,\zeta^2\}$ to obtain the 1-cochain~$H_\sigma$,
 then their coboundaries~$\delta H_\sigma$ give all six of
 the colorings as $\sigma$ ranges over
 the set of permutations of the set~$S$.
 
\enddemo

\demo{Question}  What additional physically
meaningful conservation laws are there
in other rings~$A$?  A banal example in $\Z/2\Z$ is the number of
particles modulo~2; this is an invariant since, physically, creation and
destruction of particles occurs in pairs, or, mathematically,
since all spherical diagrams in $\tilde K$ have an even number
of faces.
\enddemo

\heading{\bf Appendix 8.  Preons are determined
by charge, color, and weak isospin}
\endheading

Here we tie together all notions of this paper
and prove that the preon number (net number of $a_i$'s and $b_i$'s)
in the composition of elementary fermions of SM are determined
linearly by four of the conservation laws of SM, namely,
electric charge, weak isospin, and a color 
conservation law of Appendix~7 and its complex conjugate. 

Recall from \S 5 that electric charge is determined as $\delta h$
where $h\in C^1(K,\Q)$ is given by $h(a)=\frac 1 3, h(b)=0$.  We lift
$h$ to $\tilde K$ and multiply by 3 to get
$H\in C^2(\tilde K,\Z)$ given by $H(a_i)=1, H(b_i)=0$ for $i\pmod 3$.
The conservation law $\delta H\in B^2(\tilde K,\Z)\subset B^2(\tilde K,\C)$
will serve for our purposes the role of electric charge.

Recall also from \S 5 the weak isospin is determined by $\delta w$,
where $w\in C^1(K,\Q)$ is given by $w(a)=w(b)=\frac 1 6$,  
We lift $w$ to $\tilde K$ and multiply by 6 to get 
$W\in C^2(\tilde K,\Z)$ given by
$W(a_i)=W(b_i)=1$ for all $i\pmod 3$.  The conservation law $\delta W
\in B^2(\tilde K,\Z[\zeta]\subset B^2(\tilde K,\C)$ will serve for
our purposes the role of weak isospin.

Let $L\in C^1(\tilde K,\Z[\zeta])$ be the element of $C^1(\tilde K,\Z[\zeta])$
determined in the Example of Appendix~7, so
$L(b_i)=0$, $L(a_0)=1+\zeta^2, L(a_1)=1+\zeta,L(a_2)=\zeta+\zeta^2$.
We determined $L$ there so that $\delta L\in
B^2(\tilde K,\Z[\zeta])\subset B^2(\tilde K,\C)$ is a color conservation law.

If $\bar L$ is the complex conjugate function, then,
noting that $\bar \zeta=\zeta^2$ and $\bar \zeta^2=\zeta$,
we have
$\bar L(a_0)=1+\zeta, \bar L(a_1)=1+\zeta^2$,
$\bar L(a_2)=\zeta+\zeta^2$, and $\bar L(b_i)=0$.
Note that $\bar L$ is itself one of the color conservation
laws; in fact $\bar L=L_\sigma$, where $\sigma$ is the 
permutation of the set~$\{1, \zeta,\zeta^2\}$
that interchanges $\zeta$ and $\zeta^2$ and
leaves 1 fixed.
Note that $L+\bar L=\text{Re}(L)$, where Re denotes the
real part, so
$(L+\bar L)(a_0)=(L+\bar L)(a_1)=1$ and
$(L+\bar L)(a_2)=-2$.

\demo{Remark}  If $\sigma$ is an even permutation of
the set $\{1,\zeta,\zeta^2\}$, then $L_\sigma$ is of the
form $\zeta^jL$ for some integer~$j$, whereas 
$\bar L$ is of the form $L_\sigma$ for $\sigma$ a
transposition.
\footnote{The fact that the six color conservation laws are related
in this way is a reflection of the fact that $S_3$, the Weyl group
of the root system~$A_2$, is generated by a 3-cycle (multiplication
by $\zeta$) and a 2-cycle (complex conjugation).
}
\enddemo

We can now state the main result.

\proclaim{Theorem}  The four conservation
laws $\delta H, \delta W, \delta L$ and $\delta(L+\bar L)$
are linearly independent elements of $B^2(\tilde K,\C)$.
\endproclaim

\demo{Remark}
We calculated in Appendix~3 the dimension of the vector
space $B^2(\tilde K,\R)$ to be 4 and pointed out in the footnote
that the same result holds over every field; so it follows
from the theorem that
$B^2(\tilde K,\C)$ has basis over $\C$ given by
$\delta H, \delta W, \delta L$ and $\delta (L+\bar L)$.
The next result follows immediately from the theorem;
recall the definition of $\Delta_x$ for $x\in \{a_i,b_i\}$
from Appendix~3.

\proclaim{Corollary}  For each $x\in \{a_i,b_i\}$,
$i\pmod 3$, the conservation law~$\delta \Delta_x$
is a linear combination over $\C$ of the four conservation
laws $\delta H, \delta W, \delta L$ and $\delta (L+\bar L)$.
\endproclaim

\demo{Remark}  The significance of the corollary is that
the preon composition of all of the elementary
fermions considered in $\tilde K$ is completely
determined by conservation laws of charge, weak
isospin, and color.  Consequently the preon composition
of all composite particles (that is, the preons
occurring in the boundary labels of their van Kampen 
diagrams in $\tilde K$) is completely determined
by these four conservation laws.  So one is able to
check a hypothetical structure of a composite particle
by calculating the appropriate linear combination
of conservation laws on the constituents.  
\enddemo 

\demo{Proof of theorem}
Assume that a linear combination
$\k \delta H+\lambda \delta W +\mu \delta L+\nu \delta(L+\bar L)=0$,
where $\k, \lambda, \mu, \nu\in \C$.
Evaluate this relation on a lift of the positron $[a_0a_1a_2]$,
noting that $\d L$ and $\d \bar L$ vanish on $[a_0a_1a_2]$,
to get 
$3\kappa+3\lambda=0$.
Next evaluate the relation on a lift of the electron neutrino
$[b_0b_1b_2]$ to get
$3\lambda =0$.
It follows that $\kappa=\lambda=0$, so the
linear combination simplifies
to $\mu \delta L+\nu \delta(L+\bar L)=0$.
\enddemo

Note that $\tilde{\bar d}=[b_0b_1a_2]$
and $g\tilde{\bar d}=[b_1b_2a_0]$.
Evaluate the relation $\mu \delta L+\nu \delta(L+\bar L)=0$
on $\tilde{\bar d}$ and on $g\tilde{\bar d}$
to get the equations
$\mu L(a_i)+\nu\text{Re}L(a_i)=0$ for $i=2,0$;
that is,
$\mu(\zeta+\zeta^2)-2\nu=0$
and $\mu (1+\zeta^2)+\nu=0$.
These equations have the unique solution $\mu=\nu=0$,
so the only linear relation among 
$\d H, \d W, \d L$ and $\d (L+\bar L)$ is the trivial
relation.
It follows that they are linearly independent, and the
theorem is established.

\enddemo

The theorem of color conservation suggests that restrictions on
particles (van Kampen diagrams in $\tilde K$) can be directly
related to $\tilde K$.  With this in mind, we make two definitions.

\demo{Definition} An {\it admissible\/} particle is a
van Kampen diagram $f:D\to \tilde K$ satisfying 
\roster
\item
for each pair $\a,\a'$ of distinct 2-cells in $D$ 
similarly oriented
representing quarks with $f(\a)=f(\a')$
 there is a coloring $c\in B^2(\tilde K,\Z[\zeta])$ so
that $c(f(\a))\ne c(f(\a'))$, and
\item at most two leptons appear in $D$, and their images under $f$
are not conjugate under the covering group~$G$; that is,
if $\a$ and $\a'$ distinct similarly oriented 2-cells of $D$ which 
represent leptons, then there is no $ g\in G$ so that
$g(f(\a))=f(\a')$.
\endroster
\enddemo

Condition~(1) is independent of which conservation
law $c$ is chosen to apply in the definition.
This follows from the fact that all the 6 possible
candidates for $c$ are related to each other by
translation by $\zeta^j$ or by complex
conjugation, as we remarked earlier.  

The lepton
condition~(2) is more speculative; 
the justification is that it is difficult to
see how, for example, an electron neutrino~$\nu_e$  and a muon
neutrino~$\nu_\mu$
can bind together inside a particle.  However
 $\nu_e$ and  $\bar\nu_\mu$ together give one of
the free generators of $\pi_2(\tilde K)$ which was used
in \S 6 to give a topological explanation for neutrino
oscillations.
On the other hand, the electron~$e^-$ and electron anti-neutrino
$\bar \nu_e$ are similarly oriented in the van Kampen diagram
for the Higgs particle; but they have different preon representations
which are not conjugate under the covering group~$G$.

\demo{Definition}  An admissible particle 
$f:D\to \tilde K$ is {\it observable\/}
if the sum of the values of any cochain $c\in B^2(\tilde K,\Z[\zeta])$
on the faces $f(\a)$ is 0, where $\a$ ranges
over the 2-cells of $D$.
\enddemo

By what was remarked, if the condition is satisfied for one
color conservation law, it is satisfied for all of them.

\demo{Remark} The admissibility condition incorporates the Pauli exclusion
principle, that two bound fermions cannot have the same quantum
numbers.  The observability condition strengthens the color
conservation laws; it says no color can leak out of an observable
particle.  It is a way to formulate the experimental
observation that no isolated quark has ever been observed.
\footnote{Color can however have an indirect second order
effect on distinct particles by exchange of virtual
particles, similar to the way van der Waal forces affect
molecules even though the individual atoms are electrically
neutral.
}  
\enddemo

\demo{Ansatz
\footnote{An Ansatz is an educated guess that is
  verified later by its results.
  Source:  http://en.wikipedia.org/wiki/Ansatz}}
 All particles of SM are admissible;
 thus even virtual particles that participate in
 reactions but are not observed are admissible.  All particles
 that can be detected in an experiment 
 satisfy the observability condition.
\enddemo

\demo{Open problem} Give a consistent geometric
model that explains the algebraic conditions
in the definitions of admissibility and
observability of particles.  That is, the appearance
of the Eisenstein integers $\Z[\zeta]$ suggests that
there is a {\it discrete\/} geometry underlying these conditions, but I
do not yet know what this geometry is.
\enddemo

\heading{\bf Appendix 9. Dark Matter}
\endheading

An open problem (contrary to one of the criticisms recalled in the
Disclaimer in the final section,
 that there are no open problems) is to determine the nature of
dark matter.  Observations beginning with Zwicky in the 30's have shown
that there is more matter in galaxies than can be accounted for by observed
light sources.  I shall make a proposal here based on my model
 for what this dark matter is.  

First I list properties that dark matter must possess.
\roster
\item It must be stable over the life of a galaxy.
  Otherwise one could reconstruct it from its 
decay products.
\item It must not interact with photons, the carriers of electromagnetic
force.
\item  The only 
long-range force it can interact with is that of gravity, to account 
for those observations mentioned above.
\endroster

Observe that there is no condition on colourings, since dark matter is
by assumption not observable except for its gravitational effect.

In our model a van Kampen diagram in $\tilde K$ whose boundary label 
involves no $a$'s satisfies 2 and 3 above.  All the interior $a$'s
cancel so are not visible to outside light sources.  Since such a diagram
represents matter, it is subject to gravity.
What is not clear is that such matter can be stable.
\footnote{The van Kampen diagram (or spherical diagram)
has the same relation to a particle
as the program for a Jacquard loom has to the fabric design it creates.
That is, the van Kampen diagram is not the particle itself but is a program
for creating the particle.  So for example one of the criticisms in the
disclaimer was that the photon has no structure as far as anyone has 
observed, yet the spherical diagram for it involves the diagrams
for both the electron and positron.
The program has compound structure but the design it creates does not,
as far as anyone has determined.  In analogy 
with DNA and the cell, by now DNA is well understood, but the
machinery by which this creates a new organism, or even that of
cell division, is not presently understood.
Similarly in the Appendix 10 I create
programs for chemical elements.  This is not in conflict with the liquid
ball model consisting of floating 
protons and neutrons but rather indicates a hidden
structure, in effect a rigid crystal model that underlines this model.
}

It is unlikely that the neutrino is a candidate for dark matter.  It has
mass but it is very small (and is as yet undetermined) and propagates at nearly
the speed of light.  So it seems unlikely to be gravitationally bound
to a galaxy.

The neutron is also unlikely to be a candidate, since free neutrons are 
unstable via beta decay.  It seems unlikely that such a weak force as
gravity could stabilize otherwise free neutrons in a galaxy.

However there is an algorithmic
 process for creating from the neutron an infinite
collection of particles satisfying 2 and 3.  I do not know how to
make calculations of mass or half-lives, so I do not know whether 
any of these particles are stable over the life of a galaxy. This was
one of the basic criticisms of the Disclaimer for which I lack the
tools to answer, namely, to calculate masses. 

\demo{Construction}  Given a van Kampen diagram $D$ and an interior vertex~$v$,
we may form for each $n\ge 2$ the branched cyclic cover.  Recall this means
we remove $v$ and form the cyclic covering of degree~$n$
 of the punctured domain, then sew in a subdivided disc to obtain a new
 van Kampen diagram $D'$.  Alternatively, one can take the one-point
 compactification of the covering, and note it is a van Kampen diagram.
 
 If $D$ has edge label only involving $b$'s, then the same is true for $D'$.
Hence if $D$ satisfies 2 and 3 above, then so does $D'$
\enddemo

\demo{Example}  Let $D$ be a van Kampen diagram for the neutron;
for example, in Figure~2 interchange the letters $a$ and $b$, to get
$D$.
There is a single interior vertex, so for each $n\ge 2$ there is a
unique branched cyclic cover $D_n$ of $D$, and this satisfies 2 and 3.
I pose the open question whether any of these represent stable particles,
which is the same as asking whether all possible decay modes have higher
energy. 
\enddemo

\heading{\bf Appendix 10.  Nucleogenesis}
\endheading

I suggest here only the germ of a topological theory of origin of chemical
elements.  The models resemble those of organic chemistry.

Consider the deuteron shown below in Figure~9.   

\vskip 0.75in
 \centerline{\epsfbox[110 200 475 605]{fig9.epsi}}
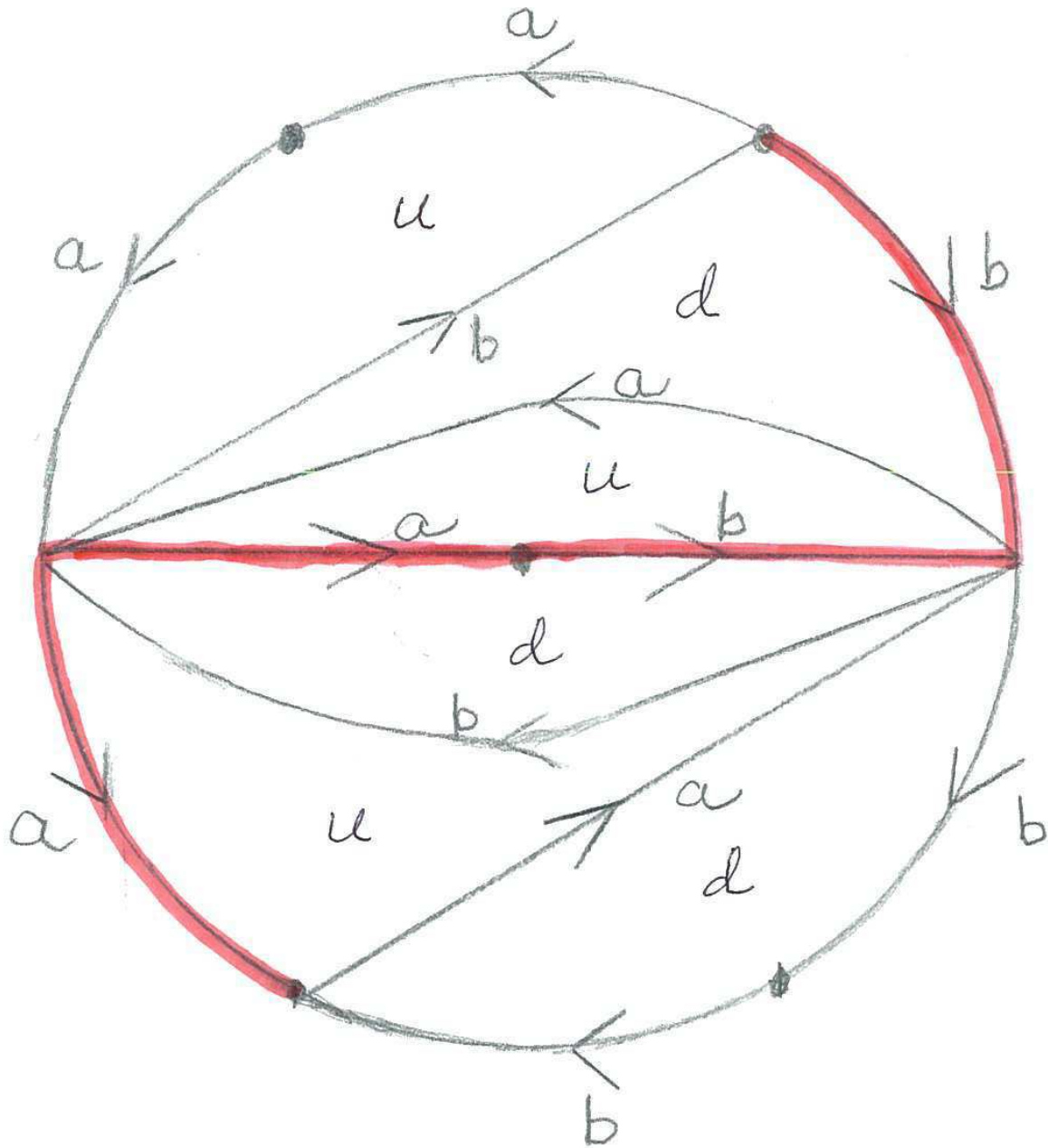
\centerline{Fig. 9}

The top part of the
diagram is the proton while the bottom is the neutron;
the up and down quarks are labeled $u$ and $d$ respectively.
  In fact two diamond moves along the red lines produce a proton and a 
neutron, so this diagram can be achieved by interactions with photons
and $Z$-bosons.  As evidence for the correctness of this model,
observe what happens when one takes branch covers about the central vertex.
The 2, 3, 5, 6, 7, 8, 10, 12, 14, 16, 18, and 20-fold cyclic
branched covers are respectively models for stable isotopes of
$_2^4\text{\rm He},\ _3^6\text{\rm Li}, \ _5^{10}\text{\rm B},
\ _6^{12}\text{\rm C},\ _7^{14}\text{\rm N},
\ _8^{16}\text{\rm O}, \ _{10}^{20}\text{\rm Ne},
\ _{12}^{24}\text{\rm Mg}, \ _{14}^{28}\text{\rm Si},
\ _{16}^{32}\text{\rm S}, \ _{18}^{36}\text{\rm Ar},
$ 
and $_{20}^{40}\text{\rm Ca}$; the last is ``the heaviest stable isotope made of
 the same number of protons and neutrons".
\footnote{Source: 
http://en.wikipedia.org/wiki/Magic$\underscore\null$number$\underscore\null$(physics)}

The upshot of this section is that stable isotopes of some of the chemical
elements can be created by diagrammatic methods for van Kampen diagrams.
I am unable to verify their stability myself for the lack of numerical methods
for associating energy to diagrams.
\footnote{But note that the energy should consist of two, possibly three,
parts.  Namely, there 
is the contribution from the 2 cells, which is the energy (or mass)
associated to the 9 fermions, which is known from experiment.
Second, there is the contribution from the edges, which should correspond to
the contribution from the strong nuclear force,
which holds the fermions together along these edges.  And third,
there is the contribution from the vertices, which must be taken into
consideration in van Kampen diagrams for the Z and W bosons and 
for the $\Delta^{++}$ resonance which are singular
disc diagrams, not topological discs.
}

There is one more point to be discussed here and that is the process of
{\bf Cloaking}.  Consider the diagram $D$ for the deuteron
in Figure 9, whose boundary edges labelled ``$a$",
 occur together as a group of three.  Take two adjacent such edges 
and connect the end points of the interval of length 2 by an edge labelled
``$b$" pointing the opposite direction.  This has the effect of attaching
(the projection from $\tilde K$ to $K$ of) an up quark to the diagram.
Then take the remaining edge labelled ``$a$" on the boundary and attach to
its endpoints two edges labelled ``$b$", that is, attach a down quark (always
referring to the projection in $K$) to the diagram.  The resulting diagram $D'$,
after attaching first the up then the down quark, is a new van Kampen diagram
containing the original deuteron as a subdiagram, but all of whose boundary
edges are labelled ``$b$".  Thus all the charges have been ``cloaked" in the 
interior and $D'$ is invisible to electromagnetic (EM) radiation.

Also observe that $D'$ has a single interior vertex.  So we may form
branched cyclic coverings of $D'$, each of which represents a particle.
So all of the stable elements constructed earlier by branched covers can
be cloaked so as to be invisible to EM radiation.
As I admitted earlier I do not have the techniques to determine whether
any of these particles are stable and hence candidates for dark matter.

In fact the cloaking process can be carried out on any diagram, so 
any such can be cloaked so as to be invisible to EM radiation.  This suggests,
although does not prove, that dark matter should be as abundant as ordinary 
matter.

\heading{Appendix 11.  The dual picture and
$\Delta^{++}$ test}
\endheading

I have worked out the consequences of the Hattori-Shupe idea in terms of
van Kampen diagrams, for which I have received considerable criticism (see
the Disclaimer), and for which I was encouraged to write an introduction
showing that this is a reasonable object to consider.  In this section I
want to point out a dual way of considering these objects which has the
advantage of being more
intuitive for physicists, and suggesting what computations must be
done to test the theory.  However in its simplest form it works
 only for van Kampen diagrams which are topological discs, and thus
fails 
for singular van Kampen diagrams, such as that for the $\Delta^{++}$ baryon
shown in Figure 7.  After presenting the basic idea, I'll take up the
modifications needed to handle singular diagrams.
The advantage of van Kampen diagrams, in addition to not being restricted 
to topological discs, is that they occur in the category of CW 2-complexes,
and thus are amenable to all the techniques of algebraic topology.
The dual pictures I shall now discuss however are a hybrid object, belonging
to graph theory with transversely oriented and labelled edges.  Thus 
convenience for mathematicians is sacrificed for intuition for physicists.  

Now for the definition of the dual picture for a van Kampen
diagram $D$.  Begin by taking the first barycentric
subdivision $D'$ of $D$ and throw away all of
$D'$ except only those edges $D'$
which join the barycenter of a face to the barycenter of an
edge which is common to two faces.  The closure of their point set union is
the dual picture $\Gamma$.  It is given the structure of
a planar graph whose vertices are the barycenters of faces
and whose edges are the union of the closures  of two edges
with a vertex (which must be the barycenter of an edge of $D$) 
in common.  
There is additional structure on $\Gamma$ in that each of 
its edges has a transverse orientation and label given by the
edge of $D$ it intersects.  

The fundamental fermions 
 are represented in $\Gamma$ by its vertices and the edges 
connecting them represent the forces acting on them. 

As an example, the dual picture for a proton (or neutron) is a planar circle
subdivided into two arcs with an edge connecting one of these vertices to 
a vertex
in the exterior.  Specifically, take the unit circle in the plane centered at
the origin with vertices $(\pm 1,0)$ together with the segment joining
$(1,0)$ to $(2,0)$.  The vertices of the circle represent up and down quarks,
and the vertex $(2,0)$ is an up quark in the case of a proton and a down
quark in the case of a neutron.  The arcs and edge represent the strong
nuclear force.
\footnote{If one knows the strength of the interaction and the
distance over which it takes place, one can estimate the
energy associated with an edge.  The mass of the quarks are
determined from the laboratory.  One must recall from an earlier
footnote that, in the case of the proton
(resp. neutron), the two up quarks (resp. down quarks) are
of different generations.  Here, to first order, one should ignore
the EM force since it is smaller than the strong nuclear force
by a factor of $10^{-3}$.
}

One can recover the original van Kampen diagram $D$ from
$\Gamma$ if $D$ is a topological disc, but this is not the case
if $D$ is singular, as is the case for the $\Delta^{++}$ baryon
represented in Fig. 5.  In addition one has lost the boundary label
of $D$.  To remedy these defects one introduces the {\it
enhanced picture\/} $\Gamma_0$.

As a graph $\Gamma_0$ is the 1-skeleton of the first
 barycentric subdivision $D'$ of the van Kampen
diagram $D$, where one discards the original edges of $D$.
  The vertices of $\Gamma_0$ are of three types, $F,E$ and $V$,
where $F$ type are barycenters of faces of $D$,
 $E$ type are barycenters of edges
and $V$ type are just the vertices of $D$.  The vertices of $F$ type
correspond to fundamental fermions (leptons or quarks).
Those of $E$ type are equipped with a transverse orientated label coming from
the label of the edge of $D$.  In particular those vertices of $E$ type
which lie on the boundary of $D$ enable one to construct the boundary
label of $D$; they are distinguished by being boundary edges of exactly
one edge of $\Gamma_0$.
In fact the entire van Kampen diagram $D$ can be reconstructed from
$\Gamma_0$ as a planar graph with its transverse labels.

\ss

\ni{\bf Example.}\footnote{I am grateful to Yong-Shi Wu for having suggested
I should consider the $\Delta^{++}$ baryon.  It turned out
to be crucial in formulating my ideas.} $\Delta^{++}$ resonance (see Fig. 5).
Here is a description of the enhanced picture $\Gamma_0$ for $\Delta^{++}$.
Let $X_1$ be the disjoint union of 6 planar edges
 where one vertex is distinguished
on each edge and called of type F.  Identify the distinguished 
vertices to get a planar graph $X$
and call the end vertices alternately in one cycle
 of type V or E, where one V type vertex is distinguished.
   Take 3 copies of $X$ and identify the
distinuished V type vertices to get the planar graph $Y$.  The images of the 
central F vertices of $X_1$ in $Y$ represent the three up quarks, and the
images of the E vertices of $X$ in $Y$ have transverse oriented labels
$a,a,b$.

The other $\Delta$-baryons have the same structure as in Fig.~5, but 
for the $\Delta^+$, $\Delta^0$, and $\Delta^{-}$ baryons the individual
quarks are changed appropriately.  They all have the
same rest mass within the error of $0.1\%$, which is what one would expect
since the dominant force holding them together is the strong nuclear force
and the electromagnetic force is smaller by a factor of a thousand.
\footnote{Source: 
https://en.wikipedia.org/wiki/Delta${\null}$\underscore$\text{baryon}$}
So my model is consistent with the data for these particles.

\ss

The point here is that if one knows the strengths of the strong interactions
and the distances between adjacent quarks (adjacent either through vertex
or through edge of the enhanced picture $\Gamma_0$) then one knows the energy 
associated to the interaction.
If one knows the rest energy of the quarks in addition, then
 one can calculate
the energy of the particle.  There is a caution here that each edge
path of length~2 in $\Gamma_0$ between two $F$ vertices represents
an interaction between the corresponding fundamental fermions.
I do not yet know how to weight these interactions if there are more than
two such paths of length~2 joining the two $F$ vertices.
However in the case of the $\Delta^{++}$ baryon there is precisely
one such path of length~2 
joining each pair of vertices of $F$ type.  By symmetry,
all the interactions they represent have the same strength and the distances
associated between the quarks are the same.  So if my correspondent who
wrote ``there are no open problems" (see Disclaimer below) is correct,
there is enough information available to calculate the rest energy of the
$\Delta^{++}$ baryon.  This amounts to a test of of the theory I have
presented.
\footnote{{\it Caveat.\/} Again I caution
that, in attempting to do the calculation for a proton,
 one must be careful to recall that the two up quarks are of different
 generations, as I showed in an earlier footnote.  In effect all calculations
 are done in $\tilde K$ and not in $K$ itself.  In the picture model
 $\Gamma$ for the proton, the two vertices on the circle are joined by
 two arcs on the circle, and one should experiment with different ways of
 weighing the strengths of the corresponding interactions, assuming that
 the theory has already survived the $\Delta^{++}$ test.
 An ``{\bf open problem}" for physicists is to make the 1-skelton ${{(\tilde K)}'}\sk 1$
  of the
 first barycentric subdivision of $\tilde K$ into a metric graph, so 
 that length of
 edges reflect strength of interactions.  Then all enhanced pictures associated
 to van Kampen diagrams will acquire pull-back lengths.
  }

\heading{Appendix 12.  Engtanglement}
\endheading

This section is independent of all that precedes and proposes
a geometrical explanation for the phenomenon of entanglement.
whereby a measurement of a particle can affect another
particle spatially separated from it (that is, the invariant
metric $-dt^2+dx^2+dy^2+dz^2$  (setting $c=1$) is positive for the two points).
I shall work in Minkowski space $\Bbb R^{1,3}$ although there is an obvious
generalization to the general pseudo-Riemannian case.  In the simplest example
of the phenomenon a particle of spin~0 located at point $P$ spontaneously
decays into two particles $X_1,X_2$ of spin $\frac12$ which fly off in
opposite directions.  I shall assume in the discussion that
both particles have non-zero rest mass, so their paths are in the
interior of the light cone at $P$.

 Conservation of angular momentum requires
that the spin measured along any axis for $X_1$ be the opposite
of that measured along the same axis for $X_2$, so if the first
is $\frac12$ along the $z$-axis, say, then the second must be
$-\frac12$.  This is true whatever the separation of $X_1$ is
from $X_2$, as long as the state of the pair remains coherent,
even if no information can pass between the particles in the
usual understanding of Minkowski space.  This prediction of quantum mechanics 
has been confirmed by experiment, so is not in dispute.  
In this example the spin of $X_1$ is not determined before
the measurement and can be an arbitrary linear combination
of ``up" and ``down" along any axis.  The measurement
picks out one of up or down, and the effect is immediate on
$X_2$: if $X_1$ is measured to be up along a given axis then $X_2$ immediately
becomes down along the same axis.
The experiments performed on Bell's inequality show that it is
not the case that $X_1$ was secretly, say, up all the time
(as might be the case if there were some hidden variables), in
which case it would follow that $X_2$ was down; that was
Einstein's explanation for the phenomenon, but Bell's inequality
(or rather its failure in quantum mechanics) shows that 
Einstein was, in this case, wrong.
\footnote{Source: 
en.wikipedia.org/wiki/Quantum$\_$entanglement
}

What
has been lacking so far has been a convincing explanation 
in the common language for
this paradoxical behavior, outside
of the formalism of quantum mechanics.  I shall offer one here,
which requires only a small modification of one's notion of
space and time, and propose an experimental test for my
explanation.

Parametrize the paths of $X_1$ and $X_2$ by proper time~$\tau$,
say, $x_i(\tau)$, $i=1,2$ and identify the points $x_1(\tau)$ and
$x_2(\tau)$ in $\Bbb R^{1,3}$ for all $\tau$ for which the wave function 
of the pair remains coherent.
The result is a singular space $S$.  No further identifications are made if
decoherence results (if, for example, a measurement is made on one particle).

One can visualize $S$ best when there is one spatial dimension~$x$ and one
time dimension~$t$, in which case the invariant form is $-dt^2+dx^2$ and
the Minkowski space is $\Bbb R^{1,1}$.
In this case $S$ is topologically a plane with a nappe of a cone resting
tangent on it.  The nappe of the cone is the image of $P$ in the $S$ and the
cone opens up in the positive $t$-direction.
\footnote{Topologically, but not analytically, this is the swallowtail
catastrophe,
en.wikipedia.org/wiki/Catastrophe$\_$theory$\#$Swallowtail$\_$catastrophe}

Let us continue the discussion in $\Bbb R^{1,1}$.
Let $Q_1$ and $Q_2$ be two points in the interior of the light cone at $P$
on opposite sides of the paths $x_1$ and $x_2$ so that $Q_2$ is on the light
cone from $Q_1$.  In $\Bbb R^{1,1}$ there is a single path for light from
$Q_1$ to reach $Q_2$.  But in the identification space $S$ there is a time-like path with shorter proper time.   Hence information
can be conveyed in $S$ between the image points by a shorter
route than that taken by light.

A more extreme example is the following.  Let $Q_1$ and $Q_2$ be chosen on
opposite sides of the cone determined by $x_1$ and $x_2$ so that the
line from $Q_1$ to $Q_2$ intersects $x_i$ at $Q_i'$, $i=1,2$.  Assume 
that the interval from $Q_1'$ to $Q_2'$ is space-like but that the
intervals from $Q_i$ to $Q_i'$ are time-like,
with the interval from $Q_1'$ to $Q_2'$ $\gg$
the sum of the aboslute values of the intervals fro $Q_i$
to $Q_i'$.  Then no information
can pass from $Q_1$ to $Q_2$ in Minkowski space.  However in $S$ the
images of $Q_i'$ are connected along a common time-like line, so in $S$
information can pass from $Q_1$ to $Q_2$. 

\demo{Remark}  It should be possible to choose $Q_1$ and 
$Q_2$ so that slight movement of a laser placed
at $Q_1$ produces interference
fringes at $Q_2$.
\enddemo 

In summary, the introduction of singularities of well-known type can
provide a geometric explanation for the paradoxical behavior of 
entanglement.

\heading{\bf Disclaimer}
\endheading

I received the following in a letter from M. Gromov:
``\dots in  order to publish the article  and not to be scorned at by the 
physicists community, Bogomolov, who  as much as  myself finds your 
article interesting and provocative, asks, if you agree, to do the 
following.
\roster
\item To have  author's (yours) disclaimer on the relation of the article  
to ``real physics" and an emphasis on the mathematical contents in it.
\item A one page appendix by a professional physicist commenting on the 
original physical input of your article. The more critical it could  be 
the better it would fare in the face of the physicists community."
\endroster

To comply with their request, I wrote the following disclaimer:

``I have been requested to include a disclaimer on the relation
of this article to ``real physics".  I am a mathematician,
not a physicist, in my seventh year of retirement; prior
to that I worked for 25 years in the field of combinatorial
group theory, and earlier than that I published articles in fields
involving algebra and topology, but not physics.
  Last November 2012
I read the Scientific American article [Li] and understood
quickly that the Harari-Shupe theory suitably interpreted
implied the solution to the generation problem, stated as an
open problem in [Li].  I read all I could find in Wikipedia, having
had no prior background in the physics involved, and asked
questions of physicists, most of which were ignored and one
of which was rebuffed with the arrogant and sarcastic (and unhelpful)
statement that ``there are no open
problems; good luck with the math." 
 The accompanying letter
from D. Singleton
shows that I was encouraged to do real physics, to calculate
masses of elementary particles and magnetic moments.
One exception was
Professor Wu, who listened to me in a meeting arranged by my
colleague Domingo Toledo and made pertinent suggestions.
He said I had solved the generation problem,
but also told me that my paper would never be accepted by
a physics journal.

As a further irony, I submitted the article to
arXiv in the category Group Theory.  The editorial board vetted
the article and resubmitted it as General Physics.
"

As for the point about criticism by a professional physicist,
I received the following letter on 1 March 2013 from
Douglas Singleton (unedited except for punctuation and TeX
formatting):

``
From: Douglas Singleton dougs\@csufresno.edu

To: Steve Gersten sg\@math.utah.edu

Hi Professor Gersten, 

Sorry for the delayed reply. I'm currently on sabbatical
 which in principle means I should 
have more time, but also in expectation of this I have taken on extra projects. 

Anyway your general question and suggestion 
(that electrons, photons, gluons, etc.) should have 
some sub-structure is interesting and in fact I tried
 to think of such a model some time ago, 
but there are a lot of problems such models face. 

First a non-important comment -- you had said that you
 have had a hard time getting physicists to look at your work. I think this 
is a combination of the fact that composite models face many challenges 
(I'll detail some below) but as 
well the formalism/notation you use is not in the standard
 tool kit of physicists so you would either need to 
``translate" your work for physicists or start a campaign 
to educate physicists to the formalism you use. 
For example there is a book on differential forms by H. Flanders from
 Purdue from the 1960s and in the foreword 
he specifically says his intent (he is a mathematician)
 was to educate physicists and engineers to the 
beauty of differential forms, wedge products, etc. In the 90s when I was in grad
 school differentials forms 
were still not part of the standard curriculum. I learned them (somewhat) on my 
own using 
Flanders book. In any case if you use unfamiliar formalism/notation this will ex
plain why most 
physicist will not take a look at your work. 

For example in the opening of your fermion paper you talk about ``Kampen diagrams
". I looked at your 
pictures and these look similar to Dynkin diagrams/root diagrams/weight diagrams
 from group theory 
(have a look at Howard Georgi's book on Lie Groups for physicists
 and your Kampen diagrams look 
similar to some of these other diagrams I mention, but then again not exactly th
e same). In turn I scanned 
both articles and did not find any obvious mention of SU(2), SU(3), U(1) or even
 string theory groups 
like E(8) x E(8) or any of the other exceptional Lie groups. 
It may be that these are discussed in a manner /formalism 
I do not know, but in any case this would explain why physicists
 have ignored your work. Also I checked the 
arXiv (both hep-th and math-ph) and there was no mention of ``Kampen diagrams" --
 I did a ``full text" 
search for ``Kampen" and found only a guy named van Kampen and then three papers 
which referenced 
van Kampen's work -- but no reference to Kampen diagrams. A google search turned
 up 
Kampen diagrams but in looking through the first 2--3 pages 
I did not see any physicists working with them. 
Again in some sense this is not really important since one can attribute this to
 the ``bad" math education of physicists, 
but as a practical matter it will mean your work will have 
an uphill battle getting attention from the physics community. 

OK but let me move on to substantive comments. The first questions you should be
 able to calculate an 
answer for if your model is correct is 

(i) What is the mass of the electron, muon, tau in terms of the more fundamental
 masses of 
your ``preons" a, b? (By the way I very much liked Harari's work
 and this inspired me to try my 
hand at this type of model building but my attempt foundered on exactly this and
 the following 
questions). Or at least you need to calculate the mass ratio e/mu and mu/tau. 

(ii) What are the values of the CKM (Cabibbo-Kobayashi-Maskawa) mixing elements 
between the 
three generations? 

If your model is correct and useful then you should be able to calculate the 
above quantities in terms of the 
input parameters of your model (I guess the parameter
 should be the masses and mixings of the a and b ``preons"). 
If you could do this then people would pay attention even if they did 
not at first understand your notation/formalism. 

Next I have some general comments which seem problematic . 

(a) Electrons, muons, quarks, etc. seem to be 
made of three ``preons" and these a, b ``preons" I assume are spin 1/2 so that the
 results electron, quarks etc. 
will come out to be spin 1/2. However if electrons are
 composite in this way (rather than being fundamental 
Dirac spinors) then their g-factor is no longer guaranteed to be g $\sim$
2 (up to QED 
corrections) but should be 
calculated from the underlying theory. In other words you need to take
 your composite electron, calculate its 
g-factor and show that it is about 2. For example the proton
 and neutron are both known to be spin 1/2 but 
they are composite spin 1/2 particles being made up of (at least at the first le
vel) quarks. Now if one 
assumes that the quarks are fundamental Dirac particles with g $\sim$ 2 
one can to a pretty good job at 
getting the g-factors for the proton and neutron
 ($\sim$ 5.59 and $\sim$ -3.83 respectively) in terms of the more fundamental 
g-factors of the quarks taken at their Dirac value.
 Note that the g-factors of the proton and neutron are *very* 
different from 2. Also the calculation of these values from QCD 
models is only at the 5\% level as compared to 
the 0.000000001\% level of QED calculations of the electron g-factor. 
This is said to be due to the very large 
non-perturbative quantum corrections coming from the strong interaction. If you 
could get your model to give the 
correct g-factor of the electron, muon, tau, etc. to the 5\% 
level this would good enough since that is what is 
done in the case of protons and neutrons. 

(b) Your photon seems to be composed of $\bar a a$ i.e.
 the photon is also a bound state. But the photon is experimentally 
known to be massless to a very high degree and 
theoretically this is said to come from gauge invariance (which can 
be broken via the Higgs mechanism but the Higgs in our Universe does not do this
 for the photon so gauge invariance means the photon 
is exactly massless and this agrees very well with experiment).Now if
 your photon is a bound state of more fundamental 
spin 1/2 particles it is highly unlikely that it would be massless 
to such a high degree. This would require an 
extremely unlikely cancellation of the binding energy of the system against the 
masses of the a and $\bar a$. In any case you need 
to show that your model gives the mass of the bound state $\bar a a$ (i.e. photon
) as zero to some fantastic accuracy. 
To do this you will need to specify the energy 
scale of the interaction which binds your ``preons" together. I also 
did not see mention of this and it is important -- what
 is the energy scale/interaction strength of the interaction which 
binds your preons together? For example in the strong 
interaction the dimensionless coupling constant at low energies 
is $\sim$ 1 -- 5 (the range is large since QCD is poorly understood in the low energy 
limit). For the E\& M interaction the 
coupling strength at low energies is $\sim$ 1/137 (which 
also explains why perturbation theory works at low energy for QED 
but fails from QCD). 

(c) Also if your photon is composed of a fermion-anti-fermion both of which are 
charged then it will have in general some 
magnetic moment, electric dipole moment, etc. It is well known 
experimentally that the photon does not have these thus 
in your model you should explain why your
 composite photon has no moments to such a high degree. The same comment 
applies to the W and Z bosons which in your composite model would
 have -- barring some cancellation -- magnetic, electric 
dipole moments in general. In fact the W from the SM has a spin and magnetic
 dipole momentum that lead to a g-factor of 
2 (same as for a fundamental Dirac particle). Because of this
 some scattering processes involving the W boson have 
``gauge amplitude zeros" in the differential scattering cross section which been 
confirmed to the 5\% level. 
The original article on this is K.O. Mikaelian, M.A. Samuel,
 and D. S ahdev,Phys. Rev. Lett. 43, 746-749(1979) . Thus your 
model would also need to how a composite W would get a g-factor so close to 2. 

Anyway the first task - if you haven't done it or if I missed 
it in the two articles - would be to calculate the electron, muon, tau 
masses (or at least their ratios) in terms of the parameters of your model. Then
 next would be to calculate the values of 
CKM elements. Another good place to test your model would be in 
predicting the mass structure of the neutrinos. The oscillation data 
for neutrinos give two possible mass structures -- regular and inverted. If your
 model could predict which is the correct mass 
structure for neutrinos this would be a good *prediction* of the model 
(the other things I ask above are post-dictions). 

Sorry for the delay and then long email. I was also going to explain a bit about
 my idea for composite leptons and 
quarks but as I said for my model I got stuck on exactly the above questions. 

Best, 

Doug "

A final word from the author, your truly.  Part of Singleton's letter refers to part II
of the article, which is under revision.  The preface to the present article is
an attempt to, at the very least, introduce the concept of van Kampen diagram,
to give a readable reference in Wikipedia, and to give a down-to-earth analogy
with Buckminster Fuller's geodetic dome, so that the concept will not appear
so strange.

\enddocument
\bye